\documentclass{aa}

\usepackage{graphicx}
\usepackage{lscape}
\usepackage{txfonts}
\usepackage[]{natbib}
\begin{document}

   \title{Searching for highly obscured AGN in the XMM-Newton serendipitous source catalog}

   \author{A. Corral
          \inst{1}
          \and
          I. Georgantopoulos\inst{1}
\and
M.G. Watson\inst{2}
\and
S.R. Rosen\inst{2}
\and
E. Koulouridis\inst{1}
\and
K.L. Page\inst{2}
\and
P. Ranalli\inst{1}
\and 
G.Lanzuisi\inst{1}
\and
G. Mountrichas\inst{1}
\and
A. Akylas\inst{1}
\and
G.C. Stewart\inst{2}
\and
J.P. Pye\inst{2}
          }

   \institute{Institute for Astronomy, Astrophysics, Space Applications, and Remote Sensing (IAASARS), National Observatory of Athens (NOA, Greece)\\
              \email{acorral@noa.gr}
\and
Department of Physics \& Astronomy, University of Leicester, Leicester, LE1 7HR, UK
             }

   \date{Received , ; accepted , }

 
  \abstract{The majority of active galactic nuclei (AGN) are obscured by large amounts of absorbing material that makes them invisible at many wavelengths. X-rays, given their penetrating power, provide the most secure way for finding these AGN. The {\it XMM-Newton} serendipitous source catalog, of which 3XMM-DR4 is the latest version, is the largest catalog of X-ray sources ever produced; it contains about half a million detections. These sources are mostly AGN. We have derived X-ray spectral fits for very many 3XMM-DR4 sources ($\gtrsim$ 114 000 observations, corresponding to $\sim$ 77 000 unique sources), which contain more than 50 source photons per detector. Here, we use a subsample of $\simeq$ 1000 AGN in the footprint of the SDSS area (covering 120 deg$^2$) with available spectroscopic redshifts. We searched for highly obscured AGN by applying an automated selection technique based on X-ray spectral analysis that is capable of efficiently selecting AGN. The selection is based on the presence of either a) flat rest-frame 
spectra from a simple power-law fit; b) flat observed spectra from an absorbed power-law fit; c) an absorption turnover, indicative of a high rest-frame column density; or d) the presence of an Fe K$\alpha$ line with a large equivalent width (EW $>$500 eV). We found 81 highly obscured candidate sources. Subsequent detailed manual spectral fits revealed that 28 of them are heavily absorbed by column densities higher than  10$^{23}$ cm$^{-2}$. Of these 28 AGN, 15 are candidate Compton-thick AGN on the basis of either a high column density, consistent within the 90\% confidence level with N$_{\rm H}$ $>$10$^{24}$ cm$^{-2}$, or a large equivalent width ($>$500 eV) of the Fe K$\alpha$ line. Another six are associated with near-Compton-thick AGN with column densities of $\sim$ 5$\times$10$^{23}$ cm$^{-2}$. A combination of selection criteria a) and c) for low-quality spectra, and a) and d) for medium- to high-quality spectra, pinpoint highly absorbed AGN with an efficiency of 80\%.}

\titlerunning{Searching for highly obscured AGN in the 3XMM-DR4 catalog}
\authorrunning{A. Corral et al.}
\maketitle

%

\section{Introduction}

The hard X-ray surveys (2-10 keV) performed with the {\it
  Chandra}\footnote{http://chandra.harvard.edu/} and {\it
  XMM-Newton}\footnote{http://xmm.esac.esa.int/} missions provide the
most unbiased census of the accretion history in the Universe because
they can penetrate large amounts of dust and gas. The deep 4Ms {\it
  Chandra} survey reached a surface density of $\sim$ 20 000 active
galactic nuclei (AGN) \rm deg$^{-2}$ (e.g., \citealt{xue}). Most X-ray
AGN are obscured by high column densities (N$_{H}$), typically above
10$^{22}$ cm$^{-2}$ (e.g., \citealt{tozzi06}, \citealt{akylas06}). In
contrast, optical quasi-stellar object (QSOs) surveys yield surface
densities lower by about two orders of magnitude because they are
prone to obscuration, although [OIII] selection diminishes the effect
of this bias \citep{bong10}. Significant improvements have been made
in mid-IR surveys, which now reach similar surface densities as X-ray
surveys (e.g., \citealt{brown06}, \citealt{delve14}) because they can
easily detect luminous QSOs. Nevertheless, they can hardly separate
obscured and, in general, low-luminosity AGN from star-forming
galaxies where the host-galaxy colors dominate the spectral energy
distribution (e.g., \citealt{barmby06},
\citealt{georgant08}). However, even the very efficient hard X-ray
surveys fail to detect a fraction of highly obscured AGN (N$_{\rm H}$
$>$ 10$^{23}$ cm$^{-2}$), and in particular, AGN with column densities
above 1.5$\times$ 10$^{24}$ cm$^{-2}$. The latter are the
Compton-thick (CT) AGN, where the primary mechanism for X-ray
attenuation is Compton scattering on electrons instead of
photoelectric absorption, which is the primary absorption mechanism at
lower column densities.\\

Several authors have suggested that the number of CT sources can be
constrained by using the spectrum of the extragalactic X-ray light,
the X-ray cosmic background (XCB). CT AGN are a basic ingredient of
XCB synthesis models (e.g., \citealt{gilli07}, \citealt{treister09},
\citealt{ballantyne11}, \citealt{akylas12}) because they are needed to
reproduce the broad peak at 20-30 keV observed in the XCB
(\citealt{frontera07}, \citealt{moretti09}). However, different
authors reported very different CT AGN fractions, the exact {\it
  intrinsic} fraction remaining uncertain by at least a factor of two,
ranging from about 10\% of the total AGN population up to 35\%. This
is most likely caused by the strong parameter degeneracy on XCB
synthesis models, as shown in \citet{akylas12}.  Owing to ultra-hard
X-ray surveys above 10 keV performed with {\it
  Swift}\footnote{http://swift.gsfc.nasa.gov/} and {\it
  INTEGRAL}\footnote{http://sci.esa.int/integral/}, CT AGN are
commonly observed in the local Universe and represent 4-20\% of local
active galaxies at energies 15-200 keV down to a flux limit of
10$^{-11}$ erg~cm$^{-2}$~s$^{-1}$ (see \citealt{burlon11} and
references therein). {\it
  NuSTAR}\footnote{http://www.nustar.caltech.edu/} is currently
extending these searches to about two orders of magnitude deeper
(\citealt{alexander13}, \citealt{lans14}).\\

Still, the identification of CT AGN in the commonly used 2-10 keV band
is difficult. Attempts to identify CT AGN have been made primarily in
deep X-ray surveys with {\it Chandra} and {\it XMM-Newton}
missions. These efforts include those of \citet{tozzi06},
\citet{comastri11} and \citet{georgant13}, all in the Chandra Deep
Field South (CDFS), the region of the sky with the deepest X-ray
observations both in {\it Chandra} and {\it XMM-Newton}. An
alternative approach is to cover a large area of the sky albeit at a
relatively bright flux limit, and at lower redshifts. According to the
XCB population synthesis models of e.g., \citet{akylas12}, the
fraction of CT sources among all AGN increases by a factor of three as
the flux limit, in the 2-10 keV band, decreases from
$\approx2\times10^{-14}$ $\rm erg~cm^{-2}s^{-1}$, the effective flux
limit of the 120 $\rm deg^2$ serendipitous XMM/SDSS survey
\citep{georg}, to 5$\times$10$^{-17}$ $\rm erg~cm^{-2}s^{-1}$, the
flux limit of the $\rm 0.12 deg^2$, 4Ms survey in the Chandra deep
field south. This means that the XMM/SDSS, composed of $\sim$ 40 000
X-ray sources detected over an area of $\sim$ 122 deg$^{2}$, contains
a factor of a couple of hundred more CT AGN than the CDFS. Given that
\citet{bright12} reported 40 CT AGN, or CT candidates, in the 4Ms CDFS
survey, this implies that there are about 8 000 CT AGN within the
XMM/SDSS survey.\\

Here, we present a fully automated selection technique of highly
obscured (N$_{H}$ $>$ 10$^{23}$ cm$^{-2}$), and CT AGN (N$_{H}$ $>$
10$^{24}$ cm$^{-2}$). The most reliable way of identifying highly
obscured AGN and CT AGN from X-ray surveys is to manually fit their
X-ray spectra to derive the actual intrinsic column density. However,
in large X-ray samples, this method is extremely time consuming, and
less reliable color-selection techniques are often used. We developed
a highly efficient selection technique (efficiency $\sim$ 80\% in
selecting highly obscured AGN), which makes use of automated spectral
fits to pinpoint this type of sources, and can be applied to large
X-ray surveys. To develop this technique, we used X-ray spectral data
from the {\it XMM-Newton} serendipitous source catalog \citep{watson},
and applied automated X-ray spectral fits implemented for the {\it
  XMM-Newton} spectral-fit database \citep{corral}. The test sample of
AGN used to develop this technique is composed of more than 1000 AGN
with spectroscopic redshifts extracted from the XMM/SDSS-DR7 (Sloan
Digital Sky Survey Data release 7) cross-correlation presented in
\citet{georg}.

\section{XMM-Newton serendipitous source catalog}
The {\it XMM-Newton} serendipitous source catalog is the largest catalog of
X-ray sources built to date \citep{watson}. In its latest release, the
3XMM-Newton {\it Data Release} 4 (hereafter 3XMM-DR4) contains
photometric information for more than 500 000 source detections
corresponding to $\sim$ 370 000 unique sources. As part of the catalog
construction, time series and spectra were also extracted if the
source counts collected in the European Photon Imaging Camera (EPIC)
were $>$ 100 (more than 120 000 detections). As a result, 3XMM-DR4
contains X-ray source and background spectra as well as ancillary
matrices for more than $\sim$ 85 000 individual sources. The
redistribution matrices used in this work are the canned matrices
provided by the {\it XMM-Newton} Science Operations Centre (SOC). Note that
the EPIC camera is composed of three detectors: one pn and two MOS
cameras\footnote{http://xmm.esac.esa.int/external/xmm\_user\_support/documentation/\\technical/EPIC/index.shtml}. The
count limit of 100 counts adopted in spectral extraction during the
construction of the 3XMM-DR4 catalog corresponds to the addition of
source (background-subtracted) counts in the three detectors. For a
detailed description of the catalog and the spectral extraction see
http://xmmssc-www.star.le.ac.uk/Catalogue/3XMM-DR4/.

\section{XMM-Newton spectral-fit database}

The {\it XMM-Newton} spectral-fit database is an European Space Agency (ESA)
funded project aimed at the construction of a catalog of automated
spectral-fitting results corresponding to all sources for which
spectral data are available within the 3XMM-DR4. The main goal is to
provide the astronomical community with a tool to construct large and
representative samples of X-ray sources according to their spectral
properties, rather than to their photometric ones. \\

The resulting spectral-fit database contains one row per source and
observation, listing source information, spectral-fit output
parameters and errors, as well as fluxes and additional information
about the goodness of fit for every model applied. A detailed
description of the database and the different spectral models applied
is presented on the project web page:
http://xraygroup.astro.noa.gr/Webpage-prodec/index.html.\\

\subsection{Automated spectral fitting}

\begin{figure}[!h]
   \centering
   \includegraphics[width=\hsize]{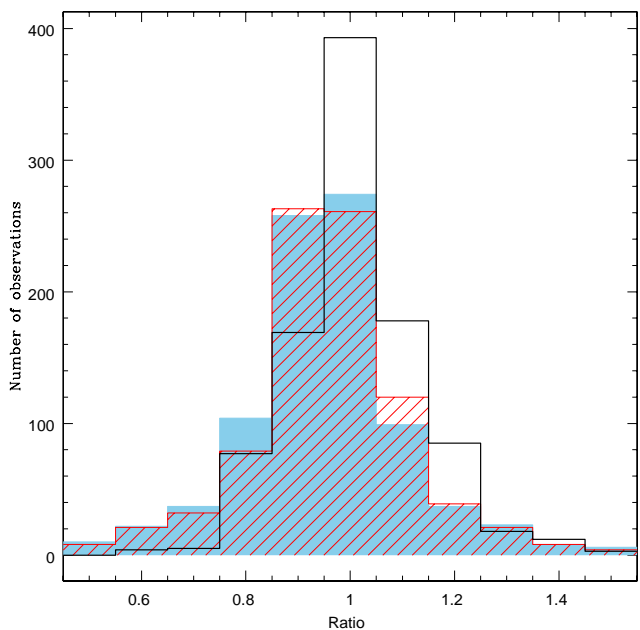}
      \caption{Distribution of normalization ratios for the three EPIC instruments: MOS2-to-MOS1 ratio (empty histogram); pn-to-MOS1 ratio (line shaded histogram); and pn-to-MOS2 ratio (filled histogram). }
         \label{norms}
   \end{figure}

The spectral-fit database is constructed by using automated spectral
fits. The software used to perform the spectral fits is {\tt XSPEC
  v12.7} (see \citealt{arnaud}), the standard package for X-ray
spectral analysis. 3XMM-DR4 source spectra are grouped to one count
per bin, and Cash statistics, implemented as C-stat in {\tt XSPEC},
are used to fit the data. This statistic was selected instead of the
more commonly used $\chi^{2}$ statistics to optimize the spectral
fitting for low-quality spectra. Grouping to one count per
bin in combination with C-stat has been proven to work very well for
low-count spectra down to 40 net (background-subtracted) counts
\citep{krumpe}. All available instruments and exposures for a single
observation of a source are fitted together. All parameters for
different instruments are tied together except for a relative
normalization, which accounts for the differences between different
flux calibrations for different EPIC instruments (MOS1, MOS2,
  and pn), which it is left free to vary. The distribution of ratios
  between the normalizations for the different instruments is shown in
  Fig.~\ref{norms}. The plotted values correspond to the sample used
  in this work, described in Sect.~\ref{testsample}, and they were
  obtained from an absorbed power-law model fit. \\

A lower limit on the number of counts in each individual spectrum was
imposed to ensure a minimum quality on the spectral fits. As a result,
not all 3XMM-DR4 spectra, but only spectra corresponding to a single
EPIC instrument, with more than 50 source counts in the full band
(0.5-10 keV) were used in the spectral fits. Parameter errors were
  computed and reported in the database at the 90\% confidence level
  ($\Delta$ C = 2.706 in one interesting parameter). The final {\it
  XMM-Newton} spectral-fit database contains spectral-fitting results
for $\gtrsim$ 114~000 detections corresponding to $\sim$ 77 000 unique
sources.\\

Three simple (absorbed power-law, absorbed thermal, and absorbed
black-body models) and three more complex (absorbed double power-law, absorbed thermal plus power-law, and absorbed black-body plus
power-law) models were implemented within the spectral-fitting
pipeline according to the number of counts in the X-ray
spectra. Simple models were applied to sources in 3XMM-DR4 with more
than 50 counts, and more complex models only to sources with more than
500 counts. These models were selected and optimized to reproduce the
most commonly observed X-ray spectral shapes among different
astronomical sources. Unlike the spectral fits used in this work (see
Sect.~\ref{autofit}), only X-ray data were used to construct the full
spectral-fit database, that is, no information about the source type or
its redshift was used during the automated spectral fits.\\

C-stat statistics lacks an estimate of the goodness of fit. To provide
a proxy of the fit quality in the spectral-fit database, goodness of
fit was estimated by using the {\tt XSPEC} command {\tt goodness}. This
command performs a number of simulations and returns the fraction of
the simulations that results in a better fit statistic. Therefore, for high return values of this command, a spectral fit with an
N\% {\tt goodness} value can be rejected at the N\% confidence
level. The reduced $\chi^{2}$ value, obtained by using C-stat as
the fitting statistic, is also included in the database.\\

Sometimes, the automated fitting process is unable to constrain all
the variable parameters during the error computation. In these cases,
spectral parameters that cannot be constrained are fixed during the
spectral fits. The fixed values of the parameters depend on the data
quality, the complexity of the model, and the energy band in which the
spectral fit is being performed. In the case of simple models, the
parameter that cannot be constrained is fixed to the value obtained by
fitting a model that only includes the corresponding model component
in the energy band that encompasses the maximum contribution of that
component. For example, if the power-law photon index cannot be
constrained in the case of the absorbed power-law model, its value is
fixed to that obtained by fitting a power-law model without absorption
in the hard (2-10 keV) band. The input parameter values for complex
models, which are also the values the parameters are fixed to if they
cannot be constrained, are the ones obtained from the spectral-fitting
results of the simple models.\\

A complete description of the XMM-newton spectral-fit database and
the automated spectral-fitting pipeline will be presented in a
forthcoming paper (Corral et al., in preparation).

\section{Automated selection of highly absorbed candidates}
\subsection{Test sample}
\label{testsample}
   \begin{figure}[!h]
   \centering
   \includegraphics[angle=-90,width=\hsize]{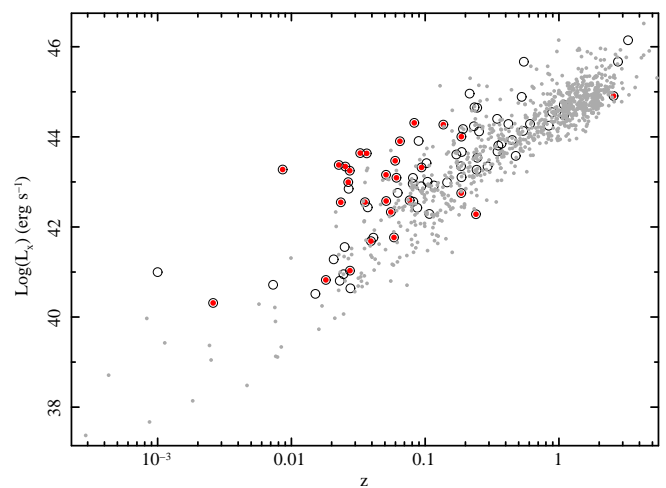}
      \caption{Intrinsic 2-10 keV luminosity versus redshift for the
        full XMM/SDSS spectroscopic sample (gray dots), the sample of
        the 81 highly absorbed candidates based on the automated
        spectral-fit selection (circles), and the 28 confirmed highly
        absorbed sources (filled circles, see text). }
         \label{lxz}
   \end{figure}
   \begin{figure}[!h]
   \centering
   \includegraphics[width=\hsize]{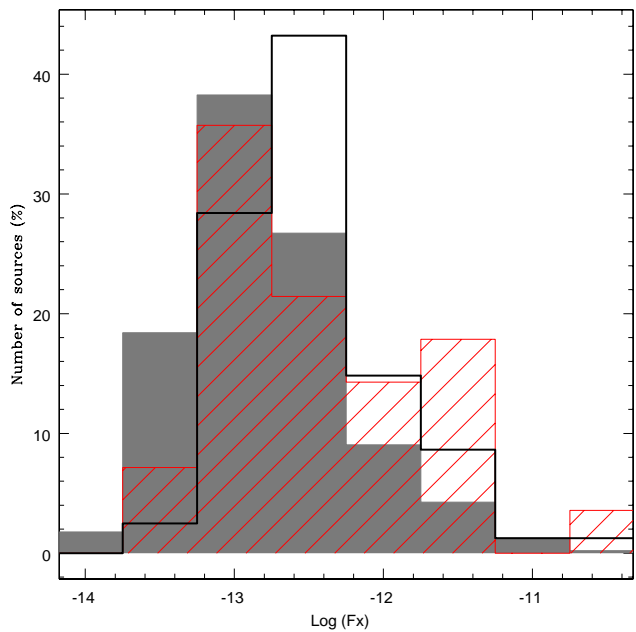}
      \caption{Observed 2-10 keV flux distribution (in c.g.s units)
        for the full XMM/SDSS spectroscopic sample (filled histogram),
        the sample of highly absorbed candidates (empty histogram),
        and the highly absorbed sources (line-shaded histogram, see
        text). }
         \label{flux}
   \end{figure}

   \begin{figure}[!h]
   \centering
   \includegraphics[width=\hsize]{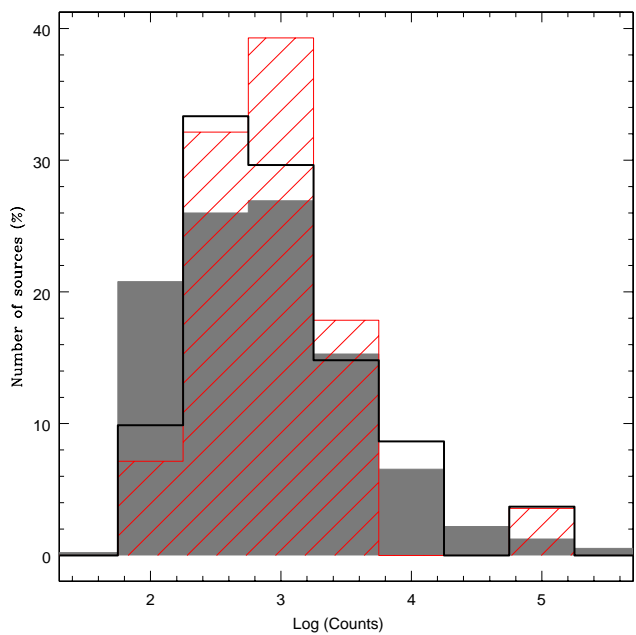}
      \caption{Distribution of total counts per source in the 0.5-10
        keV band for the full XMM/SDSS spectroscopic sample (filled
        histogram), the sample of highly absorbed candidates
        (empty histogram), and the highly absorbed sources
        (line-shaded histogram, see text). }
         \label{counts}
   \end{figure}
To test our automated selection technique, a sample of AGN was
extracted from the XMM/SDSS cross-correlation presented in
\citet{georg}. The initial sample is composed of {\bf 1015 sources}
sources, detected in the hard band (2-8 keV), for which spectral
data within the 3XMM-DR4 catalog with more than 50 counts in at
least one EPIC instrument, and spectroscopic redshifts within SDSS-DR7
were available. Figure~\ref{lxz} presents the intrinsic luminosities (in
the 2-10 keV energy band) as a function of redshift for this
sample. Observed fluxes (in 2-10 keV) and total collected counts per
source (in 0.5-10 keV) distributions are shown in Figs.~\ref{flux} and~\ref{counts}, respectively.

\subsection{Automated spectral-fit models}
\label{autofit}
The spectral models implemented for the spectral-fit database were
modified to include the effects of redshift and Galactic
absorption. Spectral-fitting results corresponding to all sources
within the XMM/SDSS-DR7 cross-correlation detected in the hard (2-8
keV) band ($\sim$ 14 000 sources) can be accessed at the spectral
database web page. Although many sources within the 3XMM-DR4 catalog
have been observed multiple times, the spectral-fitting pipeline has
been designed to fit each observation separately. Therefore, automated
spectral-fitting results are available for each individual observation
of each source.\\

For the purpose of this work, four new models were implemented by
adding a narrow Gaussian emission line to the absorbed power-law
model and the three complex models. This line is intended to
represent Fe K$\alpha$ emission line, the most commonly observed
emission line in AGN X-ray spectra. For spectra with
  fewer than 100 counts above 2 keV, the line energy and width were
  fixed to 6.4 keV, and 0.1 keV, respectively. For a larger
  number of counts, the central energy was allowed to vary within the
  6.3-6.9 keV range. Models including this emission line were
considered as complex models, and as such, they were only applied to
sources with more than 500 counts in at least one observation.

\subsection{Selection of highly absorbed candidates}

The automated spectral-fitting pipe-line, given the limited number of
spectral models that have been implemented, and the lack of
  goodness of fit, was not designed to decide which model was the
  best-fit model among those tried for each observation, but to
obtain the best possible representation of the spectral
shape. Nevertheless, acceptable fits were found for $\sim$ 80\% of the
sources in the full XMM-Newton spectral-fit database, in terms of
  {\tt goodness} values. This is not often the case for highly
  absorbed sources. This type of AGN usually displays complex X-ray
  spectra that are only poorly fitted by using the rather simple models
  used for the automated fits. However, the automated fitting results
  can be used as a proxy of the actual spectral shape, even if the
  model is an unacceptable fit, and that information can be used to
  select highly absorbed sources.\\

To obtain the most reliable selection technique, four different
automated selection criteria were explored that accommodate the spectral
characteristics most often shown by highly absorbed AGN. In all
cases, and to be able to apply our method to all the sources in the
sample with enough number of counts, automated spectral-fitting
results were used regardless of the goodness of fit for the model
under consideration. \\

A source was considered a highly absorbed candidate if its automated
spectral-fitting results fulfilled any of the following criteria (at
least in one observation, for sources with multiple
observations):
\begin{enumerate}
\item{{\bf FLATH sample (67 sources)}, flat spectrum in the hard (2-10
  keV) rest-frame band: We selected sources with a measured
  photon index in the 2-10 rest-frame band $<$ 1.4 at the 90\%
  confidence level, excluding absorption. This photon index was
  derived during the automated spectral fits as part of the absorbed
  power-law model.}
\item{{\bf FLATA sample (33 sources)}, flat spectrum in the total
  (0.5-10 keV) band: We selected sources with a measured photon
  index $<$ 1.4 at the 90\% confidence level from the absorbed
  power-law fit.}
\item{{\bf HABS sample (32 sources)}, highly absorbed sources
  (intrinsic N$_{\rm H}$ $>$ 5$\times$10$^{23}$ cm$^{-2}$) from the
  absorbed power-law fit. We did not take into account the errors
    in this case since, as pointed out before, highly absorbed sources
    are usually not acceptably fitted by a simple absorbed power-law
    model, but we can use a high column density value as an
    indication of actual heavy absorption or a complex
    spectrum. Therefore, it is important to remember that this does
    not mean that the source is actually highly absorbed, since we are
    not taking the goodness of this fit into account.}
\item{{\bf HEW sample (16 sources)}, large equivalent width sources: We
  selected spectra for which the best-fit model (the one with
  the lowest value of {\tt goodness}) includes a line with equivalent
  width (EW) $>$ 500 eV at the 90\% confidence level. Note that this
  selection criterion was only applied to sources with more
  than 500 counts collected in their X-ray spectra, {\bf 515 sources}
  in the full sample.}
\end{enumerate}

With the exception of the Fe K$\alpha$ emission line, all the other
selection criteria were chosen so as to pinpoint highly obscured
sources by using a very simple spectral model (an absorbed power law,
{\tt wabs*zwabs*pow} in {\tt XSPEC} notation), and therefore, they can
be applied to all sources within our sample with the only limitation
that the number of counts is $>$ 50 in at least one detector.\\

\begin{figure}[!h]
   \centering
   \includegraphics[angle=90,width=9cm]{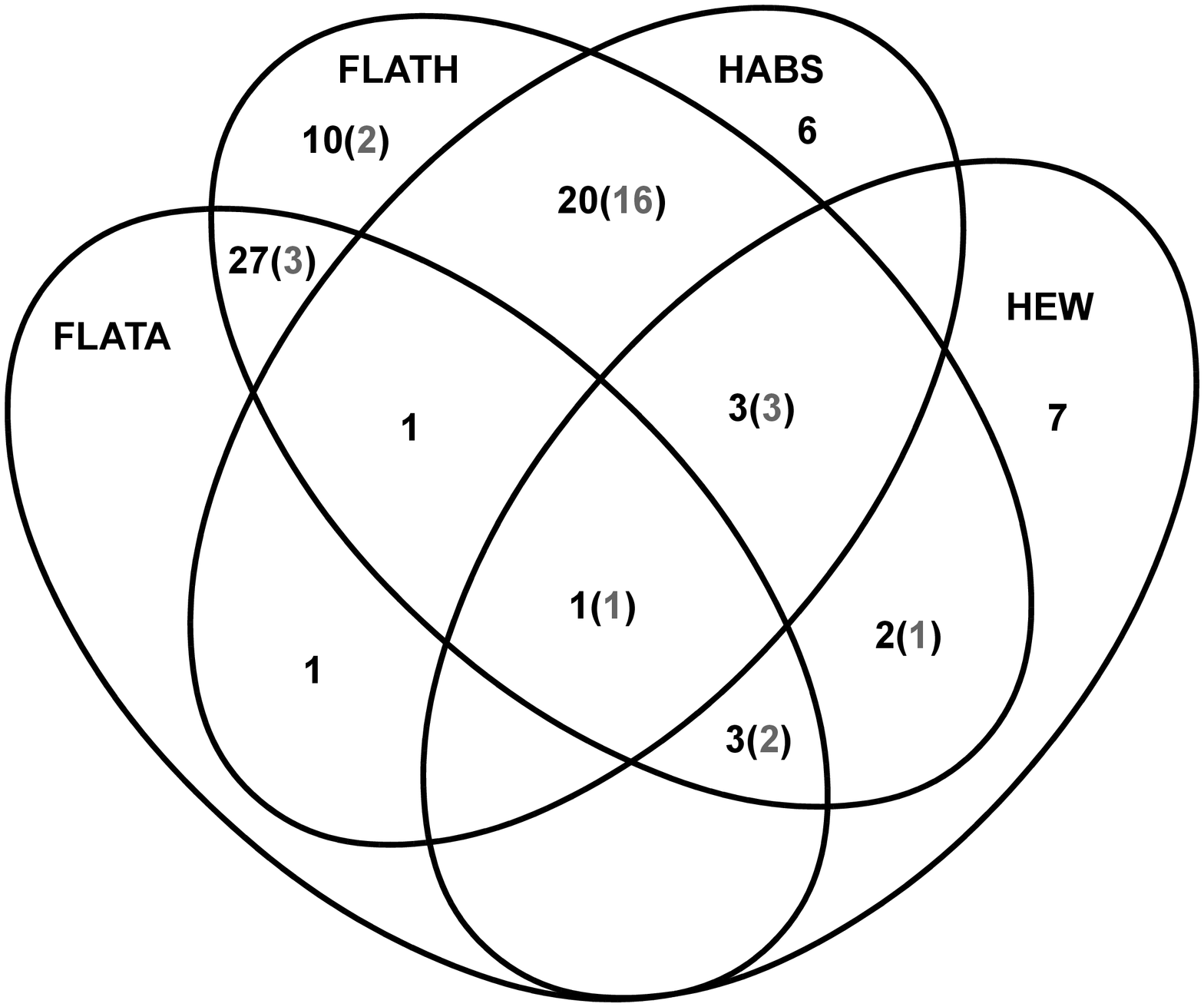}
      \caption{Venn diagram for the four samples of highly absorbed candidates. Numbers between parenthesis correspond to confirmed highly absorbed AGN according to manual spectral fits (see Sect.~\ref{manualtest}).}
         \label{venn}
   \end{figure}

Since the four samples of highly absorbed candidates have sources in
common (see Fig.~\ref{venn}), the final sample is composed of {\bf 81
  sources}. Information about the observations available in 3XMM-DR4
for these sources is listed in Table~\ref{table:1}. For sources with
multiple observations, all available observations are listed, even if
the automated spectral-fitting results classified the source as a
highly absorbed candidate in just one of the observations.

\section{Manual testing of the automated selection criteria}
\label{manualtest}
After the automated selection, all the available observations within
3XMM-DR4 of the 81 highly absorbed candidates were manually analyzed
to check if the sources were in fact highly absorbed (N$_{\rm H}$ $>$
10$^{23}$ cm$^{-2}$). To check for source variability between
different observations, the values of the photon index, intrinsic
absorption and fluxes were compared by fitting each observation
separately. Spectra for the same source and EPIC instrument
corresponding to different observations were merged if the
parameter values and fluxes were consistent within errors at the 90\%
confidence level.\\

The models applied during the manual fits are more complex than the
rather simple ones used in the construction of the database. For
example, to model the soft emission in highly oscured AGN, we tied
together both photon indices when fitting a double power-law model,
whereas they are allowed to have different values in the database. In
this way, we can separate a scattering component, the soft power-law
with the same photon index as the hard one, from an additional thermal
component(s), if present. This component(s) can become important in
highly obscured AGN at low luminosities because the contribution of
the host galaxy to the soft band becomes more and more significant.\\

The fit statistic used was C-stat, and the models applied during the
manual fits are listed below. Since targets were not excluded from
  our parent sample of 1015 AGN, many of the 81 candidates have been
  previously analyzed in detail and the results published in the
  literature, by using more complex models than those we use
  here. We restricted our spectral fits to our limited set of models to
  keep our analysis consistent. A source was considered as highly
  absorbed after the manual fits if the resulting intrinsic column
  density was higher than 10$^{23}$ cm$^{-2}$ at the 90\% confidence
  level.

\begin{itemize}
\item{PL ({\tt wabs*zwabs*pow}): photoelectrically absorbed power
  law. The absorption components wabs and zwabs (also applicable to
  the following models) correspond to absorption fixed to the
  Galactic column density at the source coordinates, and absorption
  shifted to the source redshift, respectively.}
\item{PL+L ({\tt wabs*zwabs*(pow+zgaus)}): same as PL, plus a
  narrow emission line whose central energy is fixed to 6.4 keV
  rest-frame, and its width to 0.1 keV.}
\item{WAPL ({\tt wabs*absori*pow}): same as PL, but in this case the
  redshift absorber is ionized, usually called a warm absorber.}
\item{2WAPL ({\tt wabs*absori*absori*pow}): same as WAPL, but including an additional ionized absorber.}
\item{2WAPL+L ({\tt wabs*absori*absori*(pow+zgaus)}): same as 2WAPL, plus a narrow emission line defined as in PL+L.}
\item{WAPL+L ({\tt wabs*absori*absori*(pow+zgaus)}): same as WAPL, plus a narrow emission line defined as in PL+L.}
\item{PL+T ({\tt wabs*(zwabs*pow+mekal)}: same as PL, plus a
  thermal component.}
\item{PL+T+L ({\tt wabs*(zwabs*(pow+zgaus)+mekal)}: same as
  PL+T, plus a narrow emission line defined as in PL+L.}
\item{WAPL+T+L ({\tt wabs*(absori*(pow+zgaus)+mekal)}: same as
  PL+T+L, but substituting the neutral absorber by an ionized one.}
\item{PL+R+L ({\tt wabs*(zwabs*(pow+zgaus)+pexrav)}): same as PL+R,
  plus a narrow emission line defined as in PL+L.}
\item{PL+T+R+L ({\tt wabs*(zwabs*(pow+zgaus)+pexrav)}): same
  as PL+R+L, plus a thermal component.}
\item{2PL ({\tt wabs*(zwabs*pow+pow)}): double power-law
  model in which the photon indices of both power-law components are
  tied to the same value. The scattered fraction, that is, the fraction of
  the intrinsic (power law) emission that is scattered into our line
  of sight by the absorbing material, is estimated by obtaining the
  ratio between the normalizations of both power laws.}
\item{PCPL ({\tt wabs*zpcfabs*pow}): this is in fact
  functionally the same model as 2PL, but we used this one instead if
  the scattering fraction for the 2PL model was $>$ 10\%, which
  indicates that the soft power-law component is not scattered
  emission, but transmitted emission.}
\item{PCPL+L ({\tt wabs*zpcfabs*(pow+zgaus)}): same as
  PCPL, plus a narrow emission line defined as in PL+L.}
\item{PCPL+T ({\tt wabs*zpcfabs*(pow+mekal)}): same as
  PCPL, plus a thermal component.}
\item{2PL+L ({\tt wabs*(zwabs*(pow+zgaus)+pow)}): same as
  2PL, plus a narrow emission line defined as in PL+L.}
\item{2PL+T ({\tt wabs*(zwabs*pow+mekal+pow)}): same as 2PL,
  plus a thermal component.}
\item{2PL+T+L ({\tt wabs*(zwabs*(pow+zgaus)+mekal+pow)}):
  same as 2PL+T, plus a narrow emission line defined as in PL+L.}
\item{2PL+2T+L ({\tt wabs*(zwabs*(pow+zgaus)+mekal+mekal\\+pow)}):
  same as 2PL+T+L, plus an additional thermal component.}
\end{itemize}

\begin{figure}[!h]
   \centering
   \includegraphics[width=\hsize]{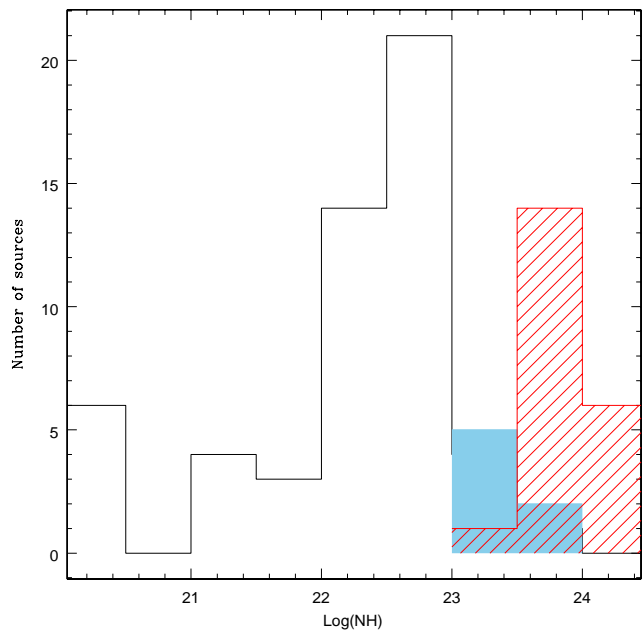}
      \caption{Column density distribution from the manual analysis
        for the 81 highly absorbed candidates: the line-shaded
        histogram correspond to the 21 CT plus near-CT AGN, filled
        histogram to the 7 highly absorbed but not CT or near-CT AGN,
        and the empty histogram to the remaining AGN.}
         \label{nhdistb}
   \end{figure}

\begin{figure}[!h]
   \centering
   \includegraphics[angle=-90,width=\hsize]{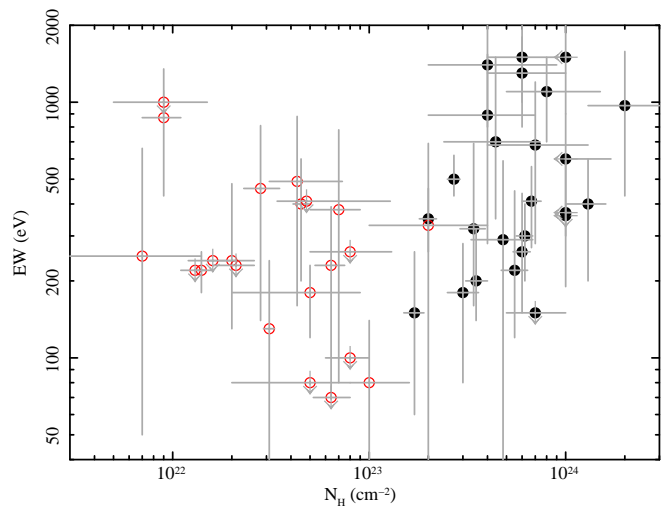}
      \caption{Column density versus Fe K$\alpha$ EW for the candidates for which the emission line is detected. Filled circles correspond to the confirmed highly absorbed AGN, empty circles to the remaining AGN.}
         \label{nhvsew}
   \end{figure}

Of the 81 highly absorbed candidates, {\bf 28 sources} display large
amounts of absorption in their X-ray spectra. The number of sources
that are best-fitted by each of the models described above are listed
in Table~\ref{sourcemodel}.  The manual spectral-fitting results,
corresponding to the 28 highly absorbed AGN, are shown in
Table~\ref{table:2}. In some cases, the addition of ionized absorption
(modeled as {\tt absori} in {\tt XSPEC}) is necessary to obtain an
acceptable fit. These sources are not considered as highly absorbed in
this analysis regardless of the column density of the ionized
absorber. In addition, if the resulting scattering fraction in a
double power-law model is $\gtrsim$ 10\%, we assumed that the model
represents partial covering absorption, which means that the soft
power-law would correspond to transmitted emission instead of
scattered emission. Therefore, these sources were not considered as
highly absorbed AGN either, again regardless of the column density of
the partial covering absorber. The manually derived column density
distribution for the 81 candidates is plotted in Fig.~\ref{nhdistb},
and the column density values versus the Fe K$\alpha$ line EW values
(in the 47 cases in which the line was detected) are plotted in
Fig.~\ref{nhvsew}. We find a similar result as reported in
\citet{fuka11}. By using {\it
  Suzaku}\footnote{http://www.isas.jaxa.jp/e/enterp/missions/suzaku/}
data, the authors reported a positive correlation between the line EW
and the measured column density, but only for high column
densities.\\

\begin{table}
\caption{\label{sourcemodel} Number of AGN manually best-fitted by each model}  
\centering          
\begin{tabular}{|l|c|c|c|c|}    
\hline\hline       
Model & Candidates & CT & near-CT & H.A.\\
(1) & (2) & (3) & (4) & (5)\\ 
\hline                    
\hline
PL & 6 & \ldots& \ldots& \ldots \\
PL+L & 8 & \ldots& \ldots& \ldots \\
WAPL & 6 & \ldots& \ldots& \ldots \\
2WAPL & 1 & \ldots& \ldots& \ldots \\
2WAPL+L & 1 & \ldots& \ldots& \ldots \\
WAPL+L & 1 & \ldots& \ldots& \ldots \\
PL+T & 5 & \ldots& \ldots& \ldots \\
PL+T+L & 1 & \ldots& \ldots& \ldots \\
WAPL+T+L & 1 & \ldots& \ldots& \ldots \\
PL+R+L & 4 & 3 & \ldots & \ldots\\
PL+T+R+L & 1 & 1 & \ldots & \ldots\\
2PL & 5 & \ldots& \ldots& 1\\
PCPL & 3 & \ldots & \ldots & \ldots\\
PCPL+L & 3 & \ldots & \ldots & \ldots\\
PCPL+T & 1 & \ldots & \ldots & \ldots\\
2PL+L & 8 & 1 & \ldots & \ldots \\
2PL+T & 3 & \ldots & 1 & 1 \\
2PL+T+L & 21 & 10 & 5 & 3\\
2PL+2T+L & 2 & \ldots & \ldots & 2 \\
\hline           
\hline       
\end{tabular}
\tablefoot{ {\it Columns:} (1) Model name (see text for details). (2) Number of AGN best-fitted by the model among the 81 highly absorbed candidates. (3) Number of CT AGN best-fitted by the model. (4) Number of near-CT AGN best-fitted by the model. (5) Number of highly absorbed AGN that are not CT or near-CT, best-fitted by the model.}
\end{table}


\begin{figure*}[h]
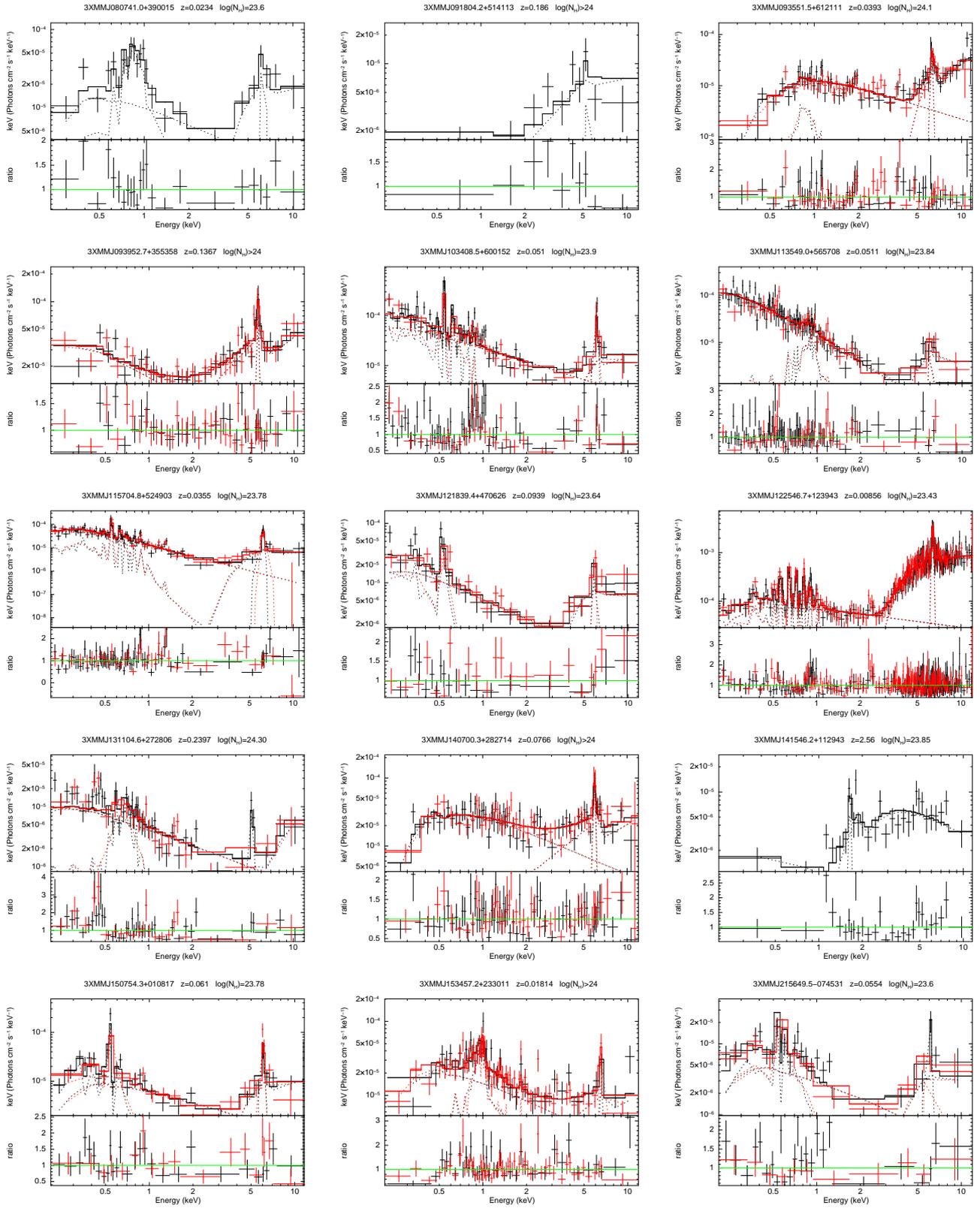

  \centering
$$
  \begin{array}{ccc}
    \includegraphics[angle=-90,width=5.5cm]{aa24129_fig81.ps} & \includegraphics[angle=-90,width=5.5cm]{aa24129_fig82.ps} & \includegraphics[angle=-90,width=5.5cm]{aa24129_fig83.ps}\\
    \includegraphics[angle=-90,width=5.5cm]{aa24129_fig84.ps} &\includegraphics[angle=-90,width=5.5cm]{aa24129_fig85.ps} & \includegraphics[angle=-90,width=5.5cm]{aa24129_fig86.ps}\\
    \includegraphics[angle=-90,width=5.5cm]{aa24129_fig87.ps} & \includegraphics[angle=-90,width=5.5cm]{aa24129_fig88.ps} & \includegraphics[angle=-90,width=5.5cm]{aa24129_fig89.ps} \\
\includegraphics[angle=-90,width=5.5cm]{aa24129_fig810.ps} & \includegraphics[angle=-90,width=5.5cm]{aa24129_fig811.ps} & \includegraphics[angle=-90,width=5.5cm]{aa24129_fig812.ps}\\
    \includegraphics[angle=-90,width=5.5cm]{aa24129_fig813.ps} & \includegraphics[angle=-90,width=5.5cm]{aa24129_fig814.ps} & \includegraphics[angle=-90,width=5.5cm]{aa24129_fig815.ps} \\
  \end{array}
$$
  \caption{Unfolded spectra and model, and data to model ratio for the
    Compton-thick candidates.}
  \label{ct}
\end{figure*}

We considered a source as a CT candidate if the value of N$_{\rm H}$ was
consistent with being higher than 10$^{24}$ cm$^{-2}$, and/or the
value of the Fe K$\alpha$ line EW was higher than 500 eV. These
sources, 15 CT candidates, are marked in boldface in
Table~\ref{table:2}, and their spectra are plotted in
Fig.~\ref{ct}. Only in two cases (3XMMJ131104.6+272806, and
3XMMJ093551.5+612111) were we able to measure a column density higher
than 10$^{24}$ cm$^{-2}$. In four cases (3XMMJ091804.2+514113,
3XMMJ093952.7+355358, 3XMMJ140700.3+282714, and 3XMMJ215649.5-074531),
we estimate that the column density is likely higher than 10$^{24}$
cm$^{-2}$ because the X-ray spectrum is reflection dominated,
meaning that there is no sign of direct emission. In five cases, the measured
column density is lower than 10$^{24}$ cm$^{-2}$, but consistent with
this value at 90\% confidence level. In the remaining four cases, the
upper limit at 90\% confidence for the column density is lower than
10$^{24}$ cm$^{-2}$, but these sources display an Fe K$\alpha$ line
with an EW $>$ 500 eV, therefore we also considered them as a CT
candidates. A high value of the Fe K$\alpha$ line EW, along with
  a very flat hard spectrum, is a characteristic of reflection-dominated spectra. We cannot exclude that the sources displaying
  high EW values are in fact reflection dominated. However, in all
  cases, the hard photon indices are either too steep or become
  steeper if they are left free to vary (in the cases in which the
  photon index is fixed to 1.9  to constrain the intrinsic
  absorption), which suggests that the hard emission is direct
  emission.  We finally classified six additional sources as near-CT
AGN. These six AGN do not fulfill our CT criteria, but their measured
column densities are very high, N$_{H}$ $\gtrsim$ 5$\times$10$^{23}$
cm$^{-2}$.\\

\begin{sidewaystable*}
 \caption{\label{table:2} Manual spectral-fitting results: Highly absorbed sources with {\rm N$_{\rm H}$ $>$ 10$^{23}$ cm$^{-2}$}}
\begin{tabular}{llccccccccc}
\hline\hline
3XMM Name(nobs) & Model & {\rm N$_{\rm H}$} & $\Gamma$ & S.F. & EW & kT & Flux & Log(Luminosity) & C-stat/d.o.f & Goodness \\
 &  & 10$^{22}$ cm$^{-2}$ &  & (\%) & eV & keV & erg cm$^{-2}$ s$^{-1}$ & erg s$^{-1}$ &  &  (\%) \\
(1) & (2) & (3) & (4) & (5) & (6) & (7) & (8) & (9) & (10) & (11) \\
\hline
\hline
 3XMMJ080535.0+240950 &  2PL+T+L & 48$^{+20}_{-15}$ & 1.9\tablefootmark{f} & 2.5 & 290$^{+300}_{-250}$ & 0.25$^{+0.20}_{-0.10}$ & 2.73$\times$10$^{-13}$ & 42.94 & 195/260 & 100 \\  
{\bf 3XMMJ080741.0+390015} & 2PL+T+L & 40$^{+30}_{-20}$ & 1.9\tablefootmark{f} & 6 & 900$^{+700}_{-600}$ & 0.7$^{+0.2}_{-0.2}$ & 2.15$\times$10$^{-13}$ & 41.94 & 202/209 & 93 \\
3XMMJ082443.2+295923 & 2PL+2T+L & 20$^{+2}_{-2}$ & 1.9$^{+0.2}_{-0.2}$ & 1 & 360$^{+130}_{-80}$ & 0.121$^{+0.017}_{-0.010}$(0.69$^{+0.06}_{-0.08}$) & 1.03$\times$10$^{-12}$ & 42.56 & 1472/1653 & 100 \\  
3XMMJ083139.0+524205(3) & 2PL+T+L & 30$^{+3}_{-3}$ & 2.3$^{+0.2}_{-0.3}$ & 1 & 180$^{+100}_{-100}$ & 0.18$^{+0.11}_{-0.08}$ & 7.71$\times$10$^{-14}$ & 42.35 & 1114/1254 & 100 \\ 
3XMMJ084002.3+294902 & 2PL+T+L & 62$^{+6}_{-6}$ & 1.9\tablefootmark{f} & 1 & 300$^{+100}_{-90}$ & 0.33$^{+0.06}_{-0.03}$ & 7.29$\times$10$^{-13}$ & 43.55 & 1103/1328 & 100 \\  
3XMMJ085331.0+175338 & 2PL+T & 50$^{+20}_{-13}$ & 1.9\tablefootmark{f} & 2 & $\ldots$ & 0.20$^{+0.10}_{-0.20}$ & 1.56$\times$10$^{-13}$ & 43.72 & 189/255 & 100 \\   
{\bf 3XMMJ091804.2+514113} & PL+R+L & $>$100 & 1.9\tablefootmark{f} & $\ldots$ & 360$^{+460}_{-360}$ & $\ldots$ & 7.49$\times$10$^{-14}$ & 42.76\tablefootmark{r} & 119/120 & 51 \\
{\bf 3XMMJ093551.5+612111} & 2PL+T+L & 130$^{+30}_{-30}$ & 1.9\tablefootmark{f} & 4 & 400$^{+200}_{-200}$ & 0.7$^{+0.4}_{-0.4}$ & 1.92$\times$10$^{-13}$ & 42.88 & 712/839 & 100 \\  
{\bf 3XMMJ093952.7+355358} & PL+R+L & $>$100 & 1.9\tablefootmark{f} & 11 & 370$^{+220}_{-180}$ & $\ldots$ & 4.38$\times$10$^{-13}$ & 43.26 & 800/904 & 100 \\  
3XMMJ095906.6+130134 & 2PL+T+L & 17$^{+2}_{-2}$ & 1.9$^{+0.3}_{-0.3}$ & 0.3 & 150$^{+110}_{-90}$ & 0.44$^{+0.15}_{-0.10}$ & 9.90$\times$10$^{-13}$ & 42.82 & 1081/1273 & 100 \\
{\bf 3XMMJ103408.5+600152} & 2PL+T+L & 80$^{+70}_{-30}$ & 1.9\tablefootmark{f} & 9 & 1100$^{+400}_{-400}$ & 0.19$^{+0.02}_{-0.02}$ & 1.99$\times$10$^{-13}$ & 42.75 & 560/533 & 66 \\  
3XMMJ104930.9+225752 & 2PL+T+L & 55$^{+8}_{-7}$ & 1.5$^{+0.3}_{-0.3}$ & 2 & 220$^{+170}_{-100}$ & 0.18$^{+0.02}_{-0.04}$ & 1.59$\times$10$^{-12}$ & 43.17 & 895/1025 & 100 \\  
3XMMJ113240.2+525701(2) & 2PL+T+L & 60$^{+7}_{-6}$ & 2.28$^{+0.13}_{-0.13}$ & 0.3 & 410$^{+150}_{-140}$ & 0.66$^{+0.09}_{-0.06}$ & 4.72$\times$10$^{-13}$ & 42.69 & 863/985 & 100 \\ 
{\bf 3XMMJ113549.0+565708} & 2PL+T+L & 70$^{+60}_{-30}$ & 2.84$^{+0.13}_{-0.13}$ & 3 & 700$^{+500}_{-400}$ & 0.8$^{+0.2}_{-0.2}$ & 5.22$\times$10$^{-14}$ & 42.43 & 515/514 & 83 \\  
3XMMJ114454.8+194635 & 2PL & 29$^{+20}_{-11}$ & 1.7$^{+0.3}_{-0.3}$ & 11 & $\ldots$ & $\ldots$ & 6.72$\times$10$^{-14}$ & 41.44 & 394/428 & 99 \\  
{\bf 3XMMJ115704.8+524903} & 2PL+T+L & 60$^{+40}_{-20}$ & 2.6$^{+0.2}_{-0.2}$ & 3 & 1500$^{+500}_{-500}$ & 0.20$^{+0.03}_{-0.02}$ & 9.26$\times$10$^{-14}$ & 42.18 & 634/617 & 74 \\  
3XMMJ120429.6+201858(2) & 2PL+T+L  & 34$^{+5}_{-5}$ & 1.8$^{+0.4}_{-0.4}$ & 1 & 320$^{+370}_{-160}$ & 0.14$^{+0.06}_{-0.05}$ & 1.73$\times$10$^{-12}$ & 42.81 & 416/477 & 100 \\  
3XMMJ120429.6+201858(1) & 2PL+T+L  & 22$^{+3}_{-3}$ & 1.8\tablefootmark{f} & 1 & 320$^{+200}_{-170}$ & 0.25$^{+0.20}_{-0.10}$ & 2.98$\times$10$^{-12}$ & 42.93 & 293/332 & 100 \\  
{\bf 3XMMJ121839.4+470626}(2) & 2PL+T+L & 44$^{+30}_{-20}$ & 2.2$^{+0.3}_{-0.4}$ & 4 & 700$^{+800}_{-350}$ & 0.14$^{+0.04}_{-0.04}$ & 7.95$\times$10$^{-14}$ & 42.80 & 412/420 & 89 \\  
{\bf 3XMMJ122546.7+123943}(1) & 2PL+T+L & 27$^{+2}_{-2}$ & 1.4$^{+0.2}_{-0.2}$ & 3 & 500$^{+120}_{-70}$ & 0.27$^{+0.03}_{-0.02}$ & 7.44$\times$10$^{-12}$ & 42.45 & 1896/2106 & 100 \\  
3XMMJ122546.7+123943(1) & 2PL+T+L & 28.7$^{+1.1}_{-1.1}$ & 1.52$^{+0.09}_{-0.09}$ & 0.7 & 220$^{+30}_{-20}$ & 0.27$^{+0.03}_{-0.01}$ & 2.11$\times$10$^{-11}$ & 42.96 & 2879/2801 & 92 \\  
3XMMJ122546.7+123943(1) & 2PL+T+L & 23.6$^{+0.4}_{-0.4}$ & 1.59$^{+0.03}_{-0.04}$ & 0.3 & 90$^{+10}_{-10}$ & 0.30$^{+0.01}_{-0.01}$ & 4.13$\times$10$^{-11}$ & 43.23 & 4268/3129 & 0 \\  
3XMMJ123843.4+092736 & 2PL+2T+L & 35$^{+5}_{-4}$ & 1.5$^{+0.3}_{-0.2}$ & 2 & 180$^{+120}_{-60}$ & 0.15$^{+0.02}_{-0.02}$(0.70$^{+0.07}_{-0.09}$) & 1.11$\times$10$^{-12}$ & 43.69 & 1530/1618 & 100 \\  
{\bf 3XMMJ131104.6+272806} & 2PL+T+L & 200$^{+120}_{-70}$ & 1.9\tablefootmark{f} & 3 & 970$^{+610}_{-540}$ & 0.19$^{+0.05}_{-0.05}$ & 3.40$\times$10$^{-14}$ & 43.80 & 398/446 & 100 \\
3XMMJ132348.4+431804 & 2PL+T+L & 60$^{+6}_{-6}$ & 2.4$^{+0.3}_{-0.2}$ & 0.2 & 260$^{+200}_{-110}$ & 0.7$^{+0.5}_{-0.3}$ & 8.70$\times$10$^{-13}$ & 43.02 & 790/971 & 100 \\  
{\bf 3XMMJ140700.3+282714} & PL+R+L & $>$100 & 1.9\tablefootmark{f} & $\ldots$ & 600$^{+300}_{-300}$ & $\ldots$ & 2.85$\times$10$^{-13}$ & 42.58\tablefootmark{r} & 646/808 & 100 \\  
{\bf 3XMMJ141546.2+112943(2)} & 2PL+L & 70$^{+30}_{-20}$ & 1.9\tablefootmark{f} & 4 & 150$^{+120}_{-150}$ & $\ldots$ & 5.97$\times$10$^{-14}$ & 45.61 & 302/365 & 100 \\  
{\bf 3XMMJ150754.3+010817} & 2PL+T+L & 60$^{+40}_{-20}$ & 1.9\tablefootmark{f} & 5 & 1300$^{+500}_{-500}$ & 0.15$^{+0.03}_{-0.04}$ & 1.12$\times$10$^{-13}$ & 42.61 & 392/426 & 99 \\  
3XMMJ150946.8+570002 & 2PL+T &  23$^{+20}_{-11}$ & 1.9\tablefootmark{f} & 3 & $\ldots$ & 0.34$^{+0.20}_{-0.07}$ & 1.36$\times$10$^{-13}$ & 39.72 & 118/124 & 83 \\  
{\bf 3XMMJ153457.2+233011(4)} & PL+T+R+L & $>$100 & 1.9\tablefootmark{f} & $\ldots$ & 1500$^{+800}_{-600}$ & 0.9$^{+0.4}_{-0.3}$ & 1.11$\times$10$^{-13}$ & 40.92\tablefootmark{r} & 556/610 & 100 \\  
{\bf 3XMMJ215649.5-074531(2)} & 2PL+T+L & 40$^{+50}_{-20}$ & 1.9\tablefootmark{f} & 9 & 1400$^{+700}_{-600}$ & 0.20$^{+0.04}_{-0.02}$ & 4.80$\times$10$^{-14}$ & 41.99 & 461/463 & 75 \\  
\hline 
\end{tabular}
\tablefoot{ {\it Columns:} (1) Source name in the 3XMM-DR4
  catalog. (2) Best-fit model. (3) Intrinsic column density. (4)
  Power-law photon index. (5) Scattered fraction in the double
  power-law model (ratio between the power-law normalizations). (6)
  Equivalent width of the Fe K$\alpha$ emission line. (7) Plasma
  temperature in the thermal model. (8) Observed flux in 2-10 keV. (9)
  Intrinsic luminosity in 2-10 keV. (10) C-stat likelihood value over
  degrees of freedom. (11) Goodness of fit estimated as 100-{\tt XSPEC
    goodness}. Numbers between parenthesis after the name of the
  source correspond to the number of merged observations. Compton-thick candidates are in boldface.
  \tablefoottext{f}{Fixed parameter.}  \tablefoottext{r}{No direct
    continuum.}  }
\end{sidewaystable*}

\section{Discussion}
\subsection{Automated selection reliability}

The automated selection method presented in this paper can be applied
to X-ray spectral data down to 50 source counts. Given the relatively
high flux limit of our test sample (typical fluxes are F$_{\rm
  X}$(2-10 keV) $\sim$ 10$^{-13}$ erg cm$^{-2}$ s$^{-1}$), most of our
absorbed sources lie at low redshifts ($<$z$>$ = 0.15), with the
exception of 3XMM141546.2+112943 at z=2.56, which is a lensed QSO, the
cloverleaf quasar H1413+117. A much more detailed analysis of this
source is presented in \citet{chartas07}, by using {\it XMM-Newton}
and {\it Chandra} data, and in which ionized absorption and disk
reflection are included in the spectral fit.  Since the only
limitation to the applicability of this method seems to be the number
of collected counts in the X-ray spectra, it could be applied to
deeper surveys, and thus, to the selection of highly obscured AGN at
higher redshifts.\\

The fraction of near-CT plus CT candidates (21 AGN with column
densities $\gtrsim$ 5$\times$ 10$^{23}$ cm$^{-2}$) in the FLATH,
FLATA, HABS, and HEW samples is 31\% (12/67), 12\% (4/33), 47\%
  (15/32), and 44\% (7/16), respectively. All the sources with
manually computed column densities higher than 10$^{23}$ cm$^{-2}$,
including the near-CT, and CT candidates, belong to the FLATH
sample. The best selection criterion is a
combination of different automated selection criteria. For all
sources, regardless of the spectral quality, the best selection
criterion is to belong to both the FLATH and the HABS samples
(FLATH+HABS subsample), 80\% of these sources are absorbed by column
densities higher than 10$^{23}$ cm$^{-2}$. For sources with more than
500 counts, that is, sources for which a selection according to the emission
line EW is possible, the best selection criterion is to belong to the
HEW and the FLATH samples (FLATH+HEW subsample). Again, 80\% of these
sources are absorbed by column densities higher than 10$^{23}$
cm$^{-2}$ according to the manual fits. Out of the 28 highly absorbed
AGN, 23 belong to either or both of these subsamples.\\

The sources within the automated selected samples that after the
manual spectral fit were found to be not highly absorbed, display one or
more of the following spectral characteristics:
\begin{itemize}
\item A reflection component, with spectral parameters consistent with
  those of unabsorbed type 1 AGN, which produces a flat photon
  index. This feature, combined with a low number of counts at high
  energies, can also mimic a high EW line.
\item Significant amount of absorption, but lower than 10$^{23}$
  cm$^{-2}$ ($\sim$ several times 10$^{22}$), totally or partially
  covering the central source.
\item Significant amount of ionized absorption. 
\item Only in some cases, we found that a low number of counts
  plus an extremely complex spectral shape classified the source in
  the HABS sample. The lower number of counts prevented us from
using
  automated complex fits, and the simple power-law fit is an extremely
  poor fit and returns a high value for the column density.
\end{itemize}

\subsection{Compton-thick candidates and previous results}

We searched the literature for previous classifications of our 28
highly absorbed sources and found a very good agreement for most cases. When previously reported intrinsic column density
or Fe K$\alpha$ EW values were available, the source was classified
according to our CT candidate definition. In three cases, a previous
estimate or measurement of the column density could not be found. Out
of the 15 sources that we found to be consistent with being a CT AGN, 12
were previously reported as CT AGN based on different techniques (see
Table~\ref{table:3}). Our classification disagrees with previous
results in only three cases: \\

\begin{itemize}
\item{3XMMJ091804.2+514113: While we classified this source as a
  reflection-dominated CT candidate, an N$_{\rm H}$ value of
  $\sim$6$\times$10$^{22}$ is reported in \citet{georg06} (although
  consistent with being highly absorbed within errors). The same
  XMM-Newton observation was used in both analyses but because
there are very few spectral counts, the differences in the
  spectral-fitting results might be due to the different spectral
  extraction. Nevertheless, our spectral data display an extremely
  flat shape that is not consistent with mild absorption.}
\item{3XMMJ093952.7+355358: If we fit the same best-fit model as in
  \citet{hard06}, that is, a double power-law plus an emission line, both
  best-fit N$_{\rm H}$ values are consistent within errors, but theirs
  is lower, probably because the second power-law photon
  index was left free to vary, which resulted in a very flat slope
  ($\Gamma$ $\sim$ 0.5). Using this model, the source would be also
  classified as a CT candidate in the current work based solely on the
  Fe K$\alpha$ line EW, but we were unable to find the relevant value in
  \citet{hard06}. Given the extremely flat hard spectrum, we classify
  this source as a reflection-dominated CT AGN in this work.}
\item{3XMMJ122546.7+123943: As for the previous source, the
  differences between our classification and that reported in
  \citet{burlon11} is caused by the EW of the iron line, which
is not reported in
  \citet{burlon11}. Moreover, this is a highly variable AGN, and a
  strong Fe K$\alpha$ line is only found in one out of the three
  observations used in this work.}
\end{itemize}

\begin{table*}
\caption{\label{table:3} Candidate Compton-thick AGN}              
\centering          
\begin{tabular}{llclc}    
\hline\hline       
Name in 3XMM-DR4 & Other names & Redshift & Previous classification & Reference\\ 
\hline                    
3XMMJ080741.0+390015 & UGC 4229 & 0.0234 & CT? & 1 \\
3XMMJ091804.2+514113 & SDSS J091804.19+514113.5 & 0.186 & ABS & 2\\
3XMMJ093551.5+612111 &  UGC 5101 & 0.0393 & CT & 3 \\
3XMMJ093952.7+355358 &  3C 223 & 0.1367 & ABS & 4\\
3XMMJ103408.5+600152 & Mrk 34 & 0.051 & CT & 5\\
3XMMJ113549.0+565708 & SBS 1133+572 & 0.0511  & CT? & 6\\
3XMMJ115704.8+524903 & 2MASX J11570483+5299036 & 0.0355 & CT & 6\\
3XMMJ121839.4+470626 & 2MASX J12183945+4706275 & 0.0939 &  CT? & 6 \\
3XMMJ122546.7+123943 & NGC 4388 & 0.00856 & HABS & 7 \\
3XMMJ131104.6+272806 &  3C 284 & 0.2397 & CT & 4\\
3XMMJ140700.3+282714 & Mrk 668 & 0.0766 & CT & 8\\
3XMMJ141546.2+112943 & H1413+117 & 2.5603 & CT & 9\\
3XMMJ150754.3+010817 & 2MASX J15075440+0108168 & 0.061 & CT? & 10\\
3XMMJ153457.2+233011 & Arp 220 & 0.01814 &  CT? & 3\\
3XMMJ215649.5-074531 & 2MASX J21564950-0745325 & 0.0554 & CT & 11 \\
\hline                  
\end{tabular}
\tablefoot{
Previous classification: ABS: 10$^{22}$ cm$^{-2}<$N$_{\rm H}<$10$^{23}$ cm$^{-2}$; HABS: N$_{\rm H}>$10$^{23}$ cm$^{-2}$; CT(?): Compton-thick
(candidate).\\
}
\tablebib{ (1) \citet{gua05}; (2) \citet{georg06}; (3) \citet{sever12}; (4) \citet{hard06}; (5) \citet{green08}; (6) \citet{lamassa09}; (7) \citet{burlon11}; (8) \citet{gua04}; (9) \citet{chartas07}; (10) \citet{lewis11}; (11) \citet{goul11}.}
\end{table*}

We also searched the literature for CT AGN within our parent sample of
1015 AGN that could have been missed as such by our automated
method. Only one AGN from the parent sample is included within the
local CT AGN reported in \citep{bright11}, NGC~3690, which was
classified as a CT AGN because of a high EW Fe K$\alpha$ emission
line. The automated analysis detected an Fe K$\alpha$ emission line
with an EW $>$ 500 eV, but consistent with being lower than that value
at the 90\% confidence level. Therefore, it was not flagged as a
highly absorbed candidate by our selection criteria. Of the sources in
out parent sample that are classified as CT AGN within the sample of
type 2 Seyfert galaxies extracted from the SDSS presented in
\citet{lamassa09}, our classification agrees in all but two cases:
SDSS~J080359.20+234520.4, and SDSS~J112301.31+470308.6. In both cases,
the automated fits return very flat photon indices in the hard band,
but because of the very low number of counts in that band, consistent
with being larger than 1.4 at the 90\% confidence level. As a result,
these two sources are not flagged as candidates by our selection
criteria. It is important to note that these two sources are
classified as CT AGN in \citet{lamassa09} based on their X-ray
luminosity to optical, and mid-IR luminosity ratios, and not from
their X-ray spectral analysis. Finally, we cross-correlated our parent
sample with the CT AGN within the two type 2 QSO samples that were
also extracted from the SDSS and reported in \citet{vignali10}, and
\citet{jia13}. Our classification agrees with that in those works
except in one case: SDSS~J091345.48+405628.2. Nevertheless, as pointed
out in \citet{jia13}, its X-ray spectrum is complex and dominated by
soft emission, and in addition, different X-ray spectral analysis of
this source, both using {\it XMM-Newton} and {\it Chandra} data, have
been published reporting a non-CT classification in some cases.

\subsection{Hardness ratios versus automated spectral-fit selection}
Rest-frame hardness ratios (or X-ray colors) have been proposed by
several other studies as an alternative to manual spectral fitting for
the selection of highly absorbed sources. The downside of these
methods is that to obtain rest-frame colors (or fluxes) from X-ray
count rates, a spectral model has to be assumed. This could strongly
decrease the accuracy of the selection technique, especially if the
assumed spectral model is a poor representation of the actual
spectral shape. To compare our proposed technique with color-selection
techniques, X-ray colors were computed for our sample following the
two different X-ray color selection techniques presented in
\citet{bright} and \citet{iwa}.\\

\begin{figure*}[!h]
  \centering
$$
  \begin{array}{cc}
    \includegraphics[angle=-90,width=7cm]{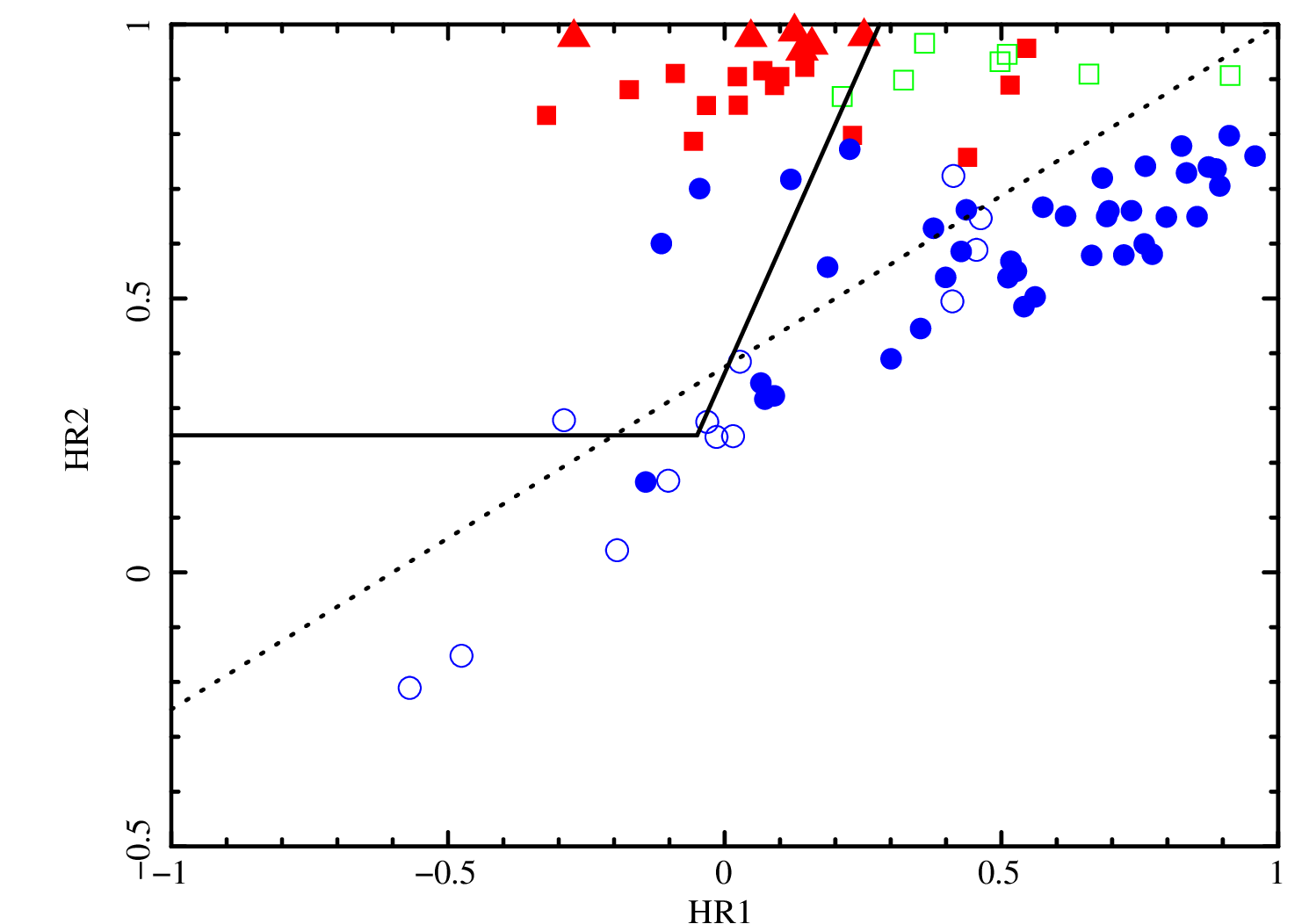} & \includegraphics[angle=-90,width=7cm]{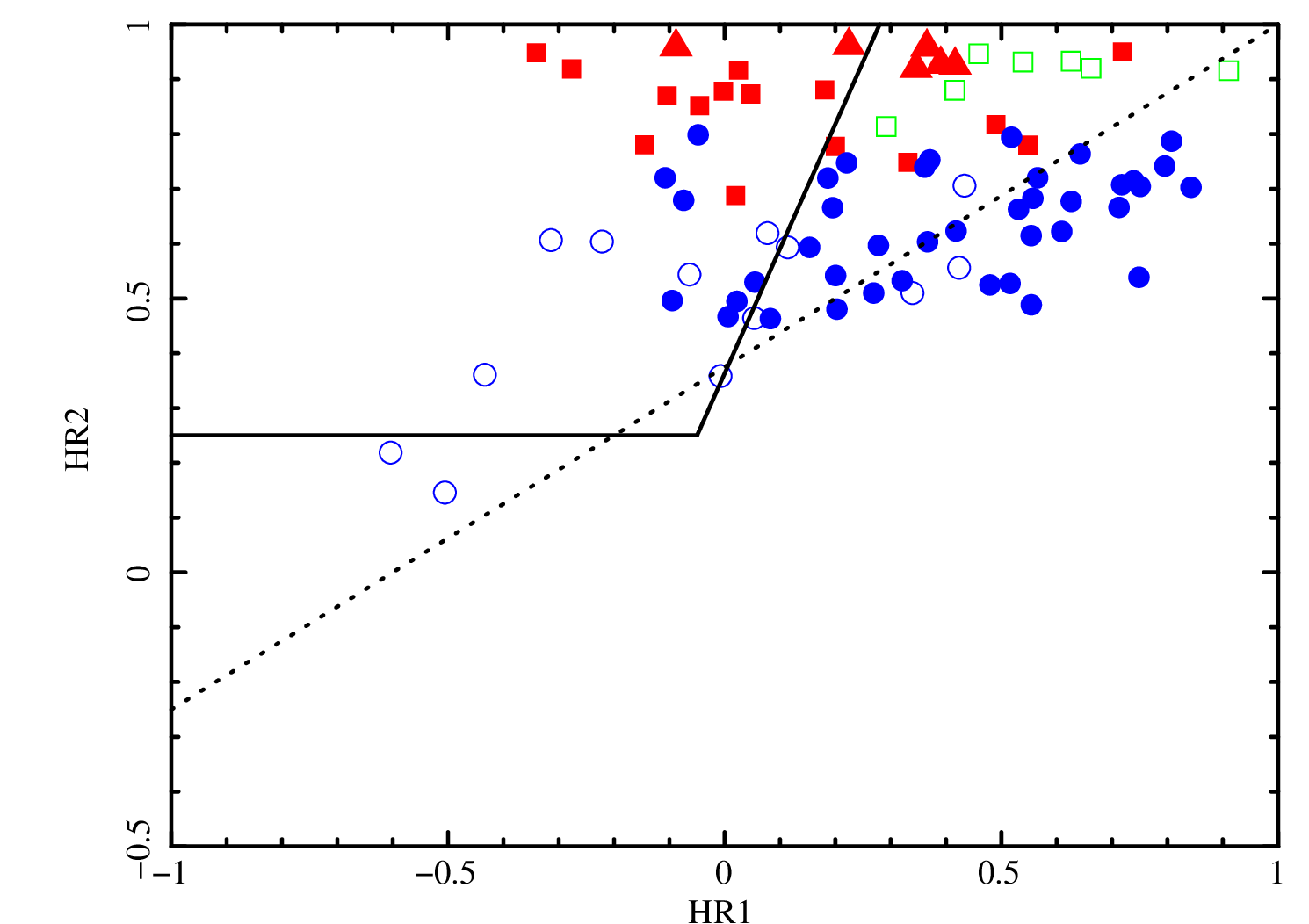}\\
    \includegraphics[angle=-90,width=7cm]{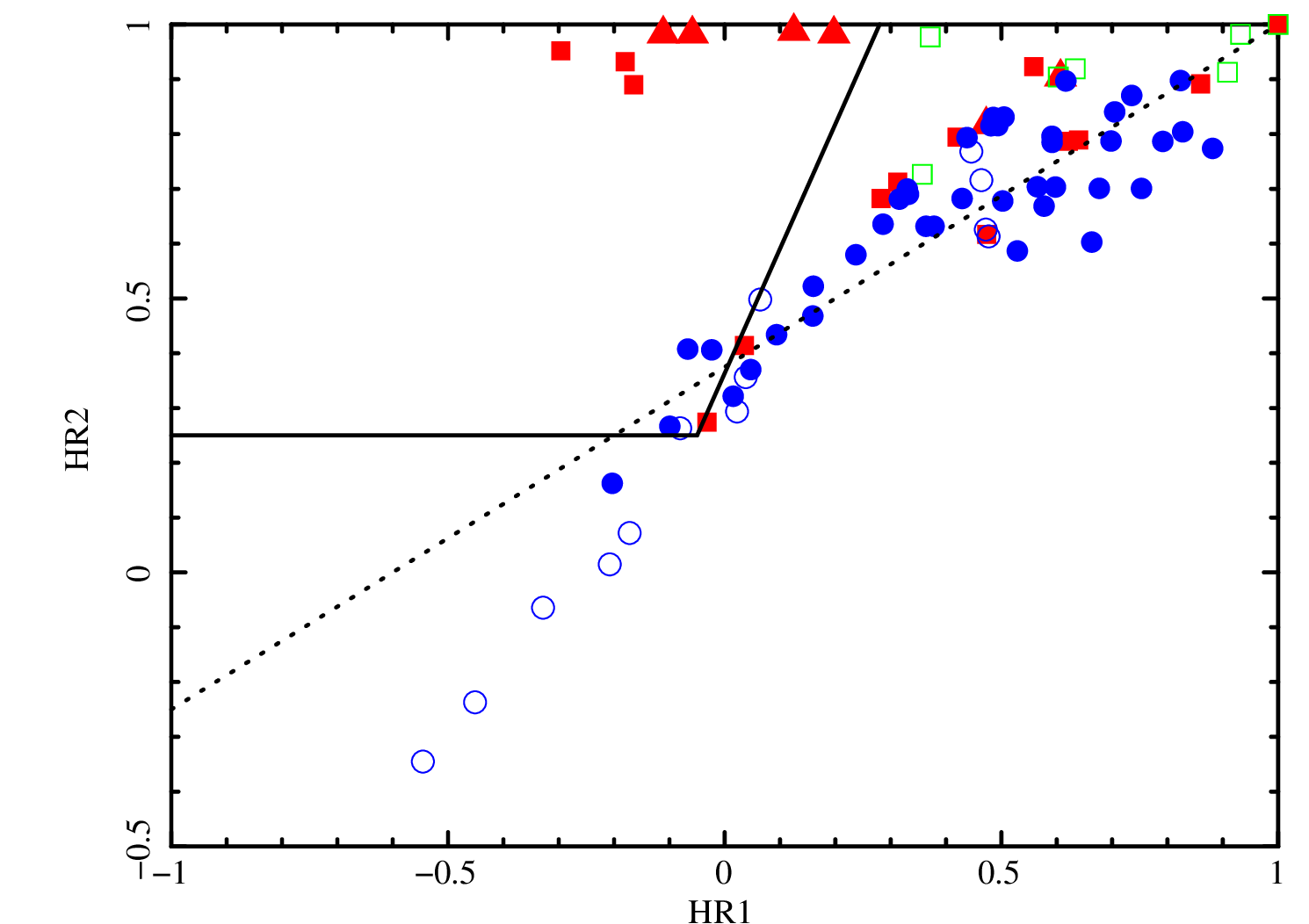} & \includegraphics[angle=-90,width=7cm]{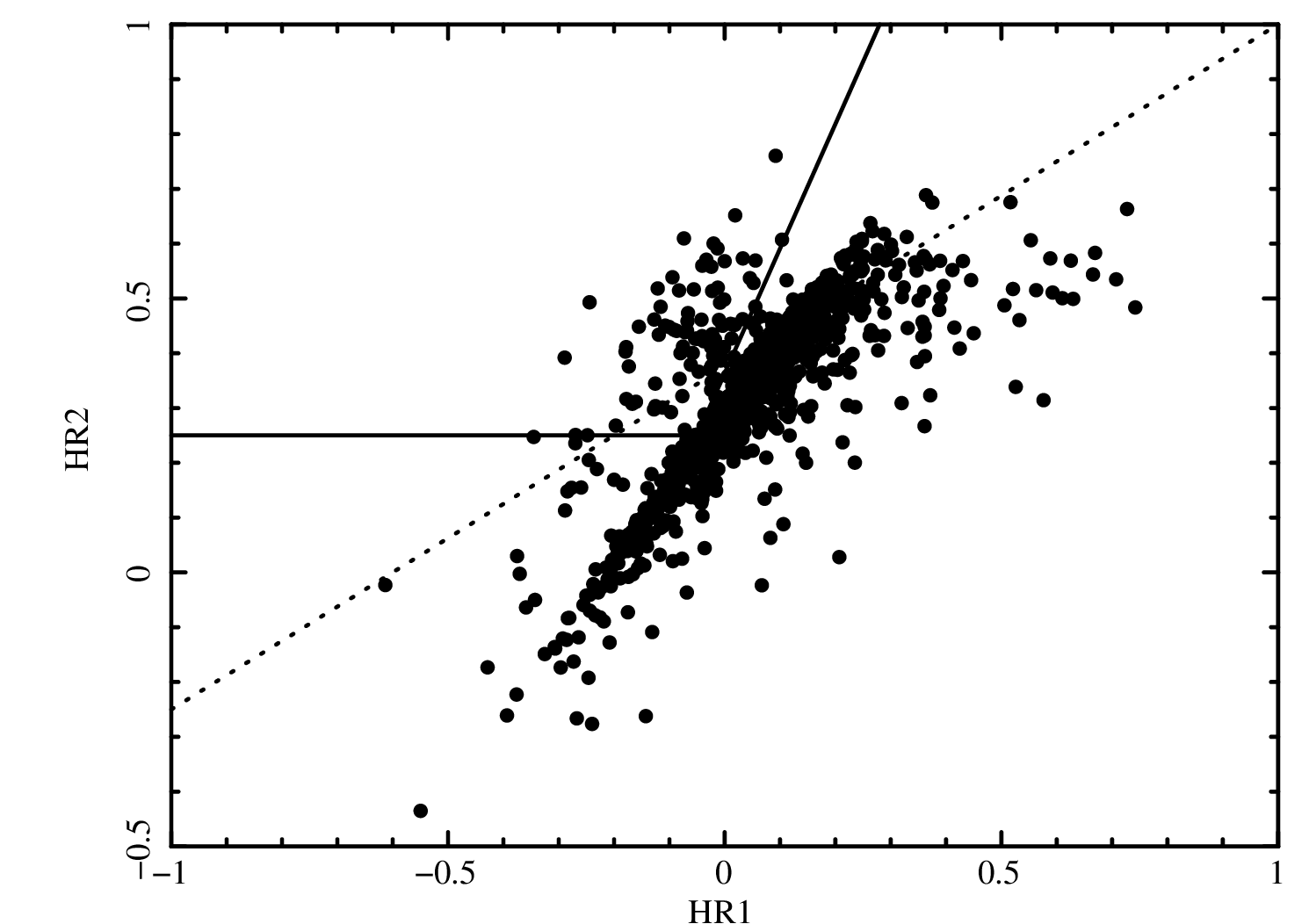}\\
  \end{array}
$$
  \caption{X-ray colors as in \cite{bright}. Different symbols correspond to column densities from manual spectral fits: CT candidate (filled squares), near-CT (filled triangles), log(N$_{H}$) $>$ 23 (empty squares), 22 $<$ log(N$_{H}$) < 23 (filled circles), and log(N$_{H}$) $<$ 22 (empty circles). Top-left: X-ray colors derived from manual spectral analysis. Top-right: colors obtained by using HEASARC PIMMS (see text for details). Bottom-left: colors obtained by using automated spectral fits. Bottom-right: colors obtained by using automated spectral fits for the sources not belonging to the sample of highly absorbed candidates.}
  \label{bcolors}
\end{figure*}

\citet{bright} presented an X-ray color selection calibrated by using
rest-frame fluxes derived from best-fit models and manual spectral
fits. To this end, they used {\it XMM-Newton} spectra for a sample of
126 local AGN, extracted from a parent sample selected in the mid-IR, for which they carried out a detailed X-ray spectral
analysis. \citet{bright} defined two X-ray colors, HR1 and HR2, based
on rest-frame fluxes computed in the rest frame bands: 1-2~keV (band
1), 2-4~keV (band 2), and 4-16~keV (band 3), as follows:
\begin{equation}
$$
{\rm HR1}=\frac{{\rm F(band 2) - F(band 1)}}{{\rm F(band 2) + F(band 1)}}\\
{\rm HR1}=\frac{{\rm F(band 3) - F(band 2)}}{{\rm F(band 3) + F(band 2)}}\\
$$
.\end{equation}
First, we applied this method to our sample of 81 highly absorbed
candidates by using the best-fit model obtained from the manual
fits. The results are plotted in Fig.~\ref{bcolors} in the top-left
panel. The dashed line corresponds to the proposed dividing line
between highly absorbed sources (N$_{\rm H}$ $>$ 10$^{23}$ cm$^{-2}$)
and mildly absorbed or unabsorbed sources in \citet{bright}. The solid
line corresponds to the wedge defined in that work to contain all
their CT AGN. We found similar results, that is, all highly absorbed
sources in our sample but one lie above the dashed line. Nevertheless,
we found a higher number of contaminants with lower N$_{\rm H}$ of
$\sim$ 20\%, while a value of only 7\% was reported in
\citet{bright}. Moreover, not all our CT candidates fall within the
solid wedge.\\

\citet{bright} also proposed that their selection technique could be
applied to X-ray colors derived by using observed count rates,
assuming a simple power-law model ($\Gamma$ = 1.4), and using
the HEASARC
portable interactive multimission software (PIMMS) to derive the
rest-frame fluxes. However, when we applied that method to obtain the
X-ray colors (Fig.~\ref{bcolors}: top-right panel), the contamination
of the highly absorbed region by sources with lower N$_{\rm H}$
increases dramatically. As a comparison, we derived the X-ray colors
by using the best-fit model from the automated spectral fits
(Fig.~\ref{bcolors}: bottom-left panel). We found a lower
contamination of the highly absorbed AGN area by unabsorbed AGN, but
still quite high. \\

As a final step, we derived X-ray colors by using automated fits for
the sources in the full sample that were not flagged as highly
absorbed candidates by the automated selection (Fig.~\ref{bcolors}:
bottom-right panel) to check whether we might be missing a significant
number of highly absorbed AGN.  We found that only 10\% of the not
selected sources lie above the dashed line. In \citet{bright}, most of
the low N$_{\rm H}$ contaminants of that region were sources with a
complex spectrum. In our case, since manual fit results are not
available for the full sample, we checked for the best-fit automated
model for these sources. In 95\% of the cases, the preferred model
was in fact a complex one. Therefore, these sources are probably not highly
absorbed sources missed by our selection criteria, but sources showing
a complex X-ray spectrum.\\

\citet{iwa} presented a different method based on observed count
rates instead of rest-frame fluxes, and designed to efficiently select
highly absorbed sources at redshifts higher than 1.7. The sample used
to test this method was composed of 47 AGN detected at high
significance in the XMM-CDFS, and with either photometric or
spectroscopic redshifts available. They also compared their results
with those obtained by using high-energy data ( $>$ 10 keV) for a
very small sample of local and well-known AGN. The method presented in
\citet{iwa} is based on the use of observed count rates in the
following rest-frame energy bands: 3-5~keV (band s), 5-9~keV (band m),
and 9-20 keV (band h). They defined two colors as s/m and h/m. Since
most of our highly absorbed candidates are at low redshift, these
three energy bands are not covered by most of our {\it XMM-Newton}
data. To be able to apply a method  as similar to that in
\citet{iwa} as possible, we used rest-frame fluxes instead of count
rates. As for the X-ray colors in \cite{bright}, a model has to be
assumed to obtain the s/m and h/m colors in this case.\\

\begin{figure*}[!h]
  \centering
$$
  \begin{array}{cc}
    \includegraphics[angle=-90,width=7cm]{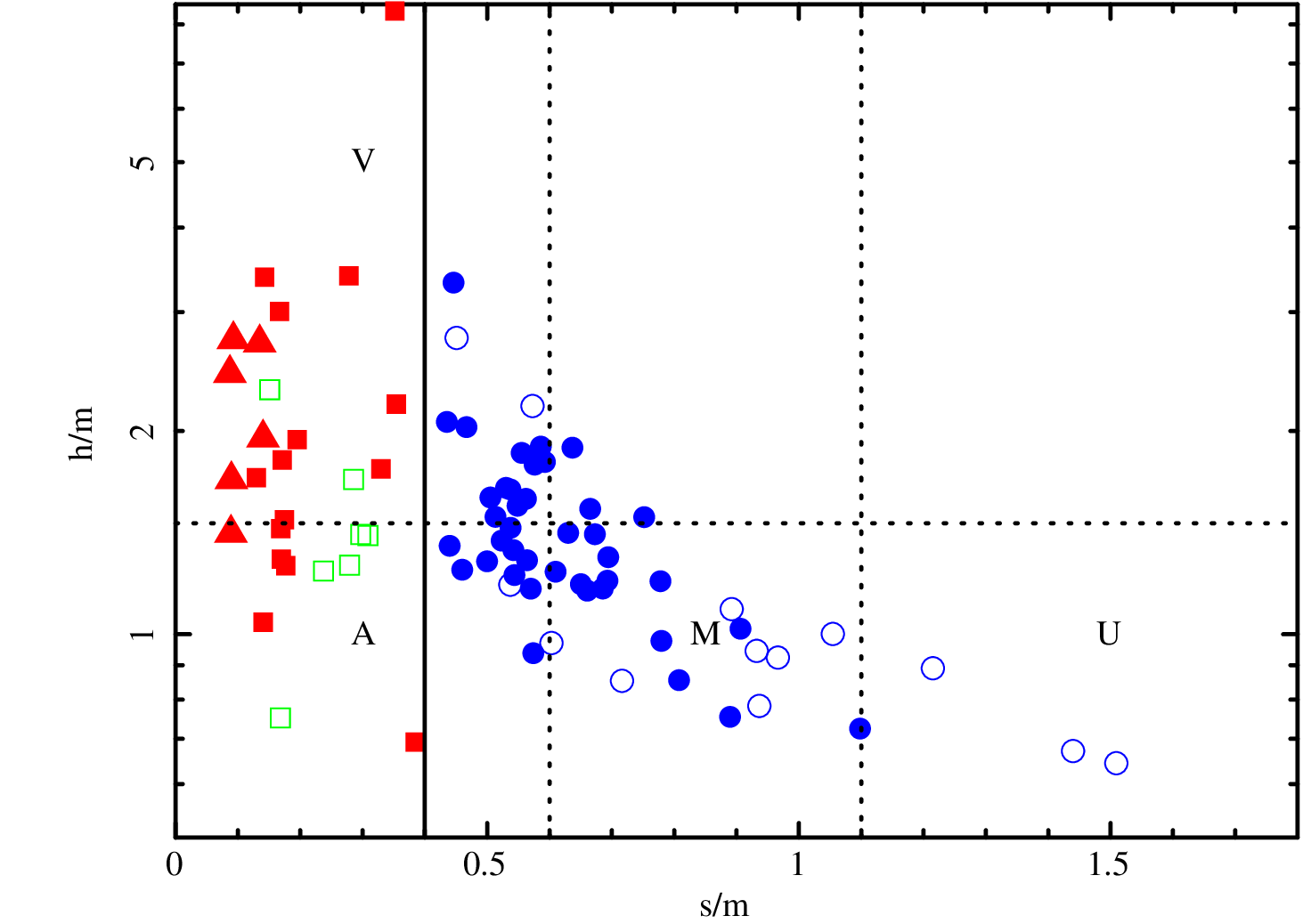} & \includegraphics[angle=-90,width=7cm]{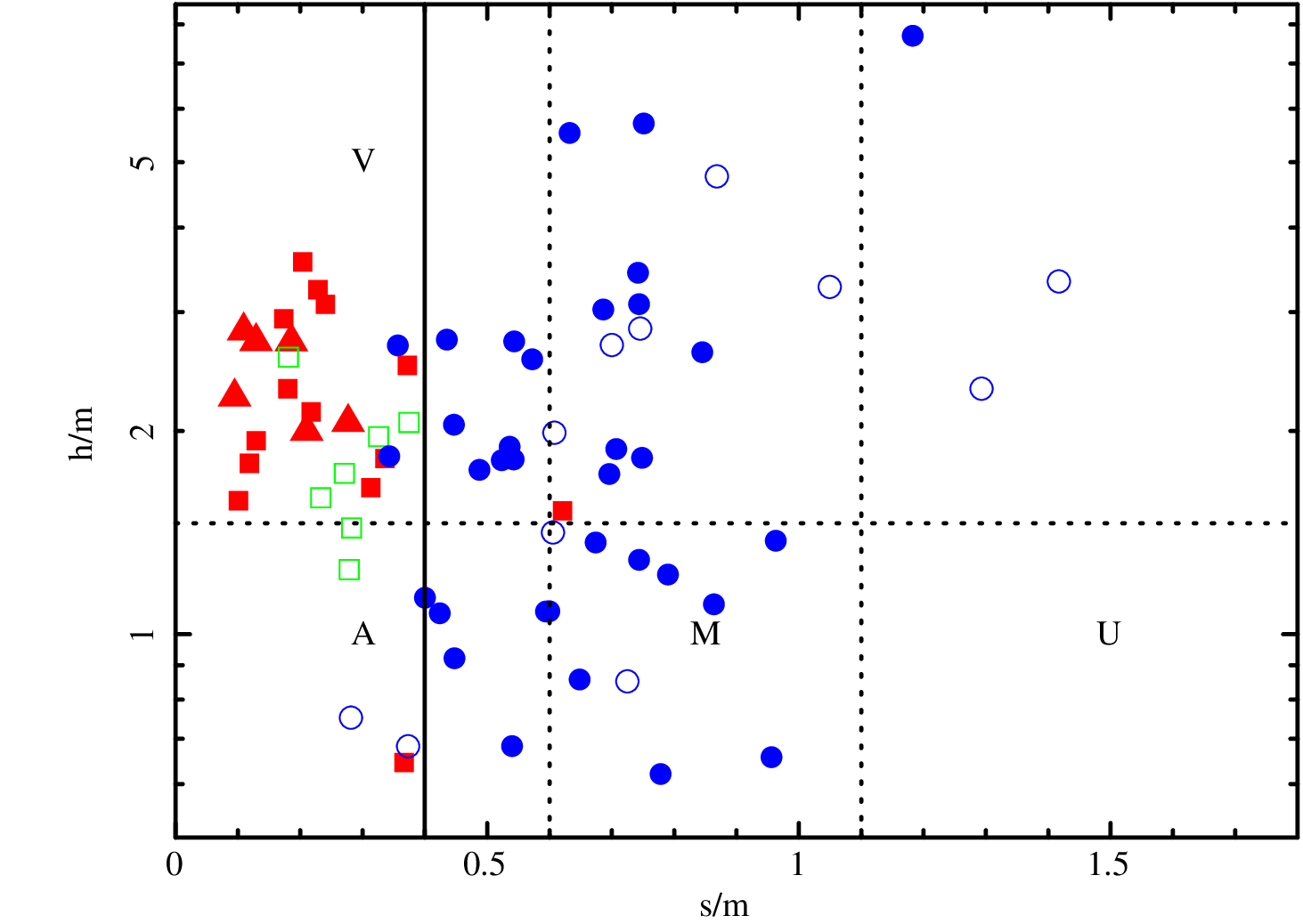}\\
    \includegraphics[angle=-90,width=7cm]{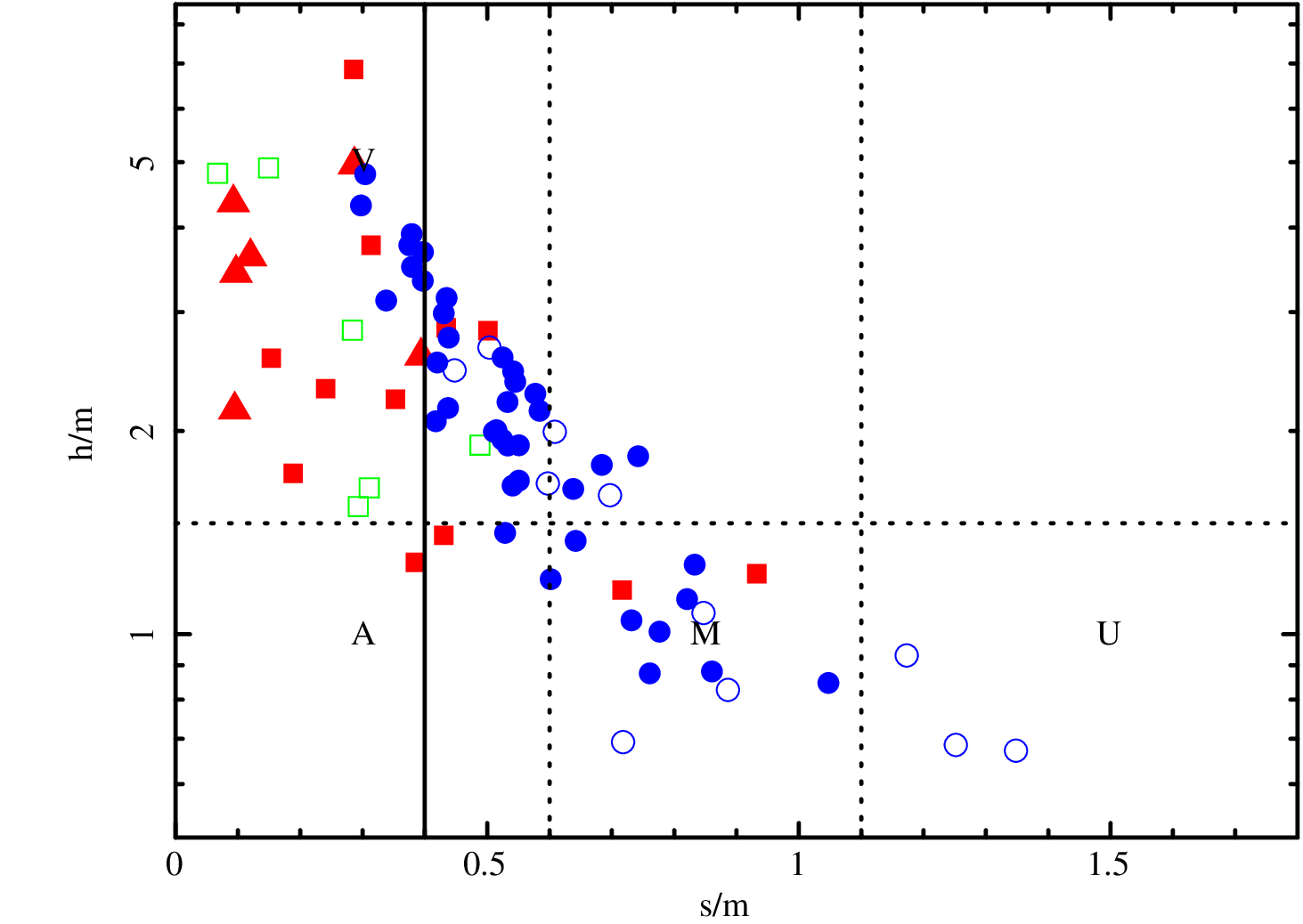} & \includegraphics[angle=-90,width=7cm]{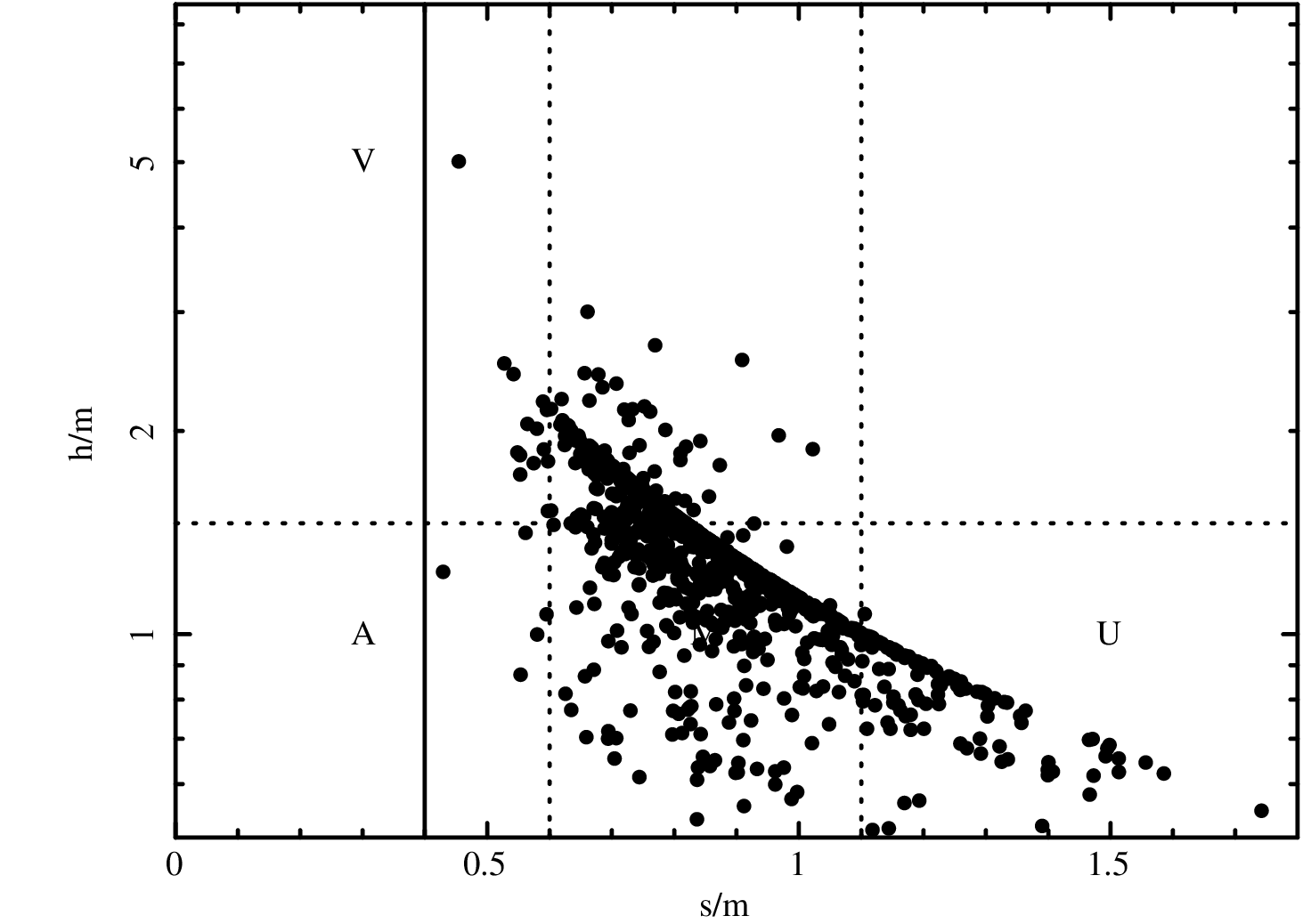}\\
  \end{array}
$$
  \caption{X-ray colors as in \cite{iwa}. Different symbols correspond to column densities from manual spectral fits: CT candidate (filled squares), near-CT (filled triangles), log(N$_{H}$) $>$ 23 (empty squares), 22 $<$ log(N$_{H}$) < 23 (filled circles), and log(N$_{H}$) $<$ 22 (empty circles). Top-left: X-ray colors derived from manual spectral analysis. Top-right: colors obtained by using HEASARC PIMMS (see text for details). Bottom-left: colors obtained by using automated spectral fits. Bottom-right: colors obtained by using automated spectral fits for the sources not belonging to the sample of highly absorbed candidates.}
  \label{icolors}
\end{figure*}

We performed the same comparison as for the colors in
\citet{bright}. The results are plotted in Fig.~\ref{icolors}. First,
we used the best-fit model from the manual fits to recover the X-ray
colors (Fig.~\ref{icolors}: top-left panel). The dashed lines limit
the different regions according to the expected column
densities. Regions U (unabsorbed), M (modestly absorbed), A
(absorbed), and V (very absorbed) correspond to sources with typical
log N$_{\rm H}$ lower than 22, 22.7, 23.4, and 23.8, respectively. We
drew a tentative limit (solid line) that separates sources with
N$_{\rm H}$ $>$ 10$^{23}$cm$^{-2}$ in our sample. This
method very efficiently separates moderate or unabsorbed AGN from
highly absorbed AGN. However, as for the \cite{bright}, and the
automated selection criteria, we cannot separate highly absorbed
sources from CT candidates. We repeated the same exercise by using
PIMMS (assuming a power-law model with photon index equal to 1.4,
Fig.~\ref{icolors}: top-right panel), and by using the automated
spectral-fitting results (Fig.~\ref{icolors}: bottom-left panel). The contamination of low N$_{\rm H}$ sources is similar in
both cases. In the automated fits, highly obscured AGN
occupy a broader region of the plot, while for PIMMS
colors, the highly obscured region is contaminated by unabsorbed
AGN. The method of \citet{iwa} seems to separate AGN with
different amounts of intrinsic absorption much better than that
presented in \citet{bright}.\\

Finally, we applied this method to the full sample by using the
best-fit model from the automated fits (Fig.~\ref{icolors}:
bottom-right panel). All the sources not classified as
highly absorbed candidates by our automated criteria fall outside the
highly absorbed AGN region.

\section{Conclusions}

We have derived X-ray spectral fits for very many 3XMM-DR4
sources ($\sim$ 77 000) that contain more than 50 photons per detector
\citep{corral}. Here, we used a subsample of $\simeq$ 1000 AGN in the
common SDSS and 3XMM area (covering 120 deg$^2$) with spectroscopic
redshifts available.  We searched for highly obscured AGN by applying an
automated selection technique based on an automated X-ray spectral
analysis. In particular, the selection was based on the presence of a)
flat spectra with a photon index lower than 1.4 at the 90\% confidence
level in the 2-10 keV rest-frame spectra, b) flat spectra with a photon
index lower than 1.4 at the 90\% confidence level in the 0.5-10 keV
observed spectra, c) an absorption turnover, indicative of a high
rest-frame column density, or d) an Fe K$\alpha$ line
with a large equivalent width ($>$500 eV). We found 81 candidate highly
obscured sources. Subsequent detailed manual spectral fits revealed that
28 are heavily obscured with a column density of N$_H$ $>$ 10$^{23}$
cm$^{-2}$. Of these 28 sources, six are near-CT AGN with a column
density of N$_H$ $\sim$ 5$\times$ 10$^{23}$ cm$^{-2}$. Finally, 15 are
candidate CT AGN on the basis of either a high column density,
consistent within the 90\% confidence level with 10$^{24}$ cm$^{-2}$,
or a large equivalent width ($>$500 eV) of the Fe K$\alpha$ line.\\

Our automated method is very efficient in selecting highly
absorbed AGN (N$_H$ $>$ 10$^{23}$ cm$^{-2}$), with a successful rate
of 80\%:
\begin{itemize}
\item For low-quality spectra, and by using only results from a simple
  absorbed power-law fit, 80\% (20 out of 25 AGN)
  of the sources whose X-ray spectra were flagged as flat in the 2-10
  rest-frame band and for which the automatically derived column
  density was higher than 5$\times$10$^{23}$ cm$^{-2}$ were highly absorbed, as tested by using manual spectral fits.
\item For medium- to high-quality spectra, 80\% (7 out of 9 AGN) of
  the sources with an automatically detected high EW Fe K$\alpha$
  line, plus a flat continuum in the 2-10 keV rest frame band were
  highly absorbed, as tested by using manual spectral fits.
\end{itemize}

We compared our results with rest-frame color CT AGN selection
techniques developed by \citet{bright} and \citet{iwa}. The method of \citet{iwa}, modified by using a spectral model to
obtain rest-frame fluxes, was the best for separating
highly absorbed (N$_H$ $>$ 10$^{23}$ cm$^{-2}$) from
moderately to unabsorbed sources.
   
\begin{acknowledgements}
      Based on observations obtained with {\it XMM-Newton}, an ESA
      science mission with instruments and contributions directly
      funded by ESA Member States and NASA. A. Corral acknowledges
      financial support by the European Space Agency (ESA) under the
      PRODEX program. P. Ranalli and E. Koulouridis acknowledge
      financial support from the 'Support to Postdoctoral Researchers'
      projects PE9-3493 and PE9-1145 which are jointly funded by the
      European Union and the Greek Government in the framework of the
      programme {\it Education and lifelong learning}.  G. Mountrichas
      and G. Lanzuisi acknowledge financial support from the THALES
      project 383549, which is jointly funded by the European Union
      and the Greek Government in the framework of the programme {\it
        Education and lifelong learning}. We thank the referee for
      providing constructive comments and suggestions that
      helped to improve this paper.
\end{acknowledgements}

\bibliographystyle{aa}
\bibliography{aa2014_24129}

%
%
%
\Online

\onllongtab{
\begin{landscape}
\begin{longtable}{lllrccccc}
\caption{\label{table:1} Sample of highly absorbed candidates.}\\
\hline
\hline
{\rm 3XMM Name} & Redshift & Observation & Counts & HEW & FLATH & HABS & FLATA & CAND \\
 (1) & (2) & (3) & (4) & (5) & (6) & (7) & (8) & (9) \\
\hline
\endfirsthead
\caption{continued.}\\
\hline
3XMM Name & Redshift & Observation & Counts & HEW & FLATH & HABS & FLATA & CAND \\
 (1) & (2) & (3) & (4) & (5) & (6) & (7) & (8) & (9) \\
\hline
\endhead
\hline
\endfoot
\hline
\endlastfoot
3XMMJ021047.0-100153 & 0.5401 & 0204340201 & 162 & N & Y & N & Y & Y\\
\hline
3XMMJ073534.9+435414 & 0.1918 & 0083000101 & 245 & N & Y & Y & N & Y\\
\hline
3XMMJ080020.9+263648 & 0.0268 & 0504101201 & 17553 & N & Y & N & Y & Y\\
\hline
3XMMJ080535.0+240950 & 0.0597 & 0203280201 & 220 & N & Y & Y & N & Y\\
\hline
3XMMJ080741.0+390015 & 0.0234 & 0138951401 & 217 & N & Y & Y & N & Y\\
\hline
3XMMJ082035.6+210404 & 0.0151 & 0505930301 & 4685 & Y & N & N & N & Y\\
 &  & 0108860501 & 3131 & N & N & N & N & N\\
\hline
3XMMJ082443.2+295923 & 0.0253 & 0504102001 & 3224 & N & Y & Y & N & Y\\
\hline
3XMMJ083139.0+524205 & 0.0585 & 0092800201 & 654 & N & Y & N & Y & Y\\
 & & 0502220201 & 325 & N & Y & N & Y & Y\\
 & & 0502220301 & 475 & N & Y & N & Y & Y\\
\hline
3XMMJ083924.9+575231 & 0.187 & 0406541201 & 142 & N & N & N & N & N\\
 &  & 0406541001 & 64 & N & N & N & N & N\\
 &  & 0406540201 & 758 & Y & N & N & N & Y\\
\hline
3XMMJ084002.3+294902 & 0.0648 & 0504120101 & 2088 & N & Y & Y & N & Y\\
\hline
3XMMJ085331.0+175338 & 0.1866 & 0305480301 & 189 & N & Y & Y & N & Y\\
\hline
3XMMJ091127.5+055053 & 2.7633 & 0083240201 & 791 & N & Y & N & N & Y\\
\hline
3XMMJ091804.2+514113 & 0.186 & 0084230601 & 77 & N & Y & N & N & Y\\
\hline
3XMMJ091828.5+513929 & 0.1855 & 0084230601 & 395 & N & Y & N & Y & Y\\
\hline
3XMMJ092718.3+304538 & 0.5267 & 0200730101 & 126 & N & N & Y & N & Y\\
\hline
3XMMJ093255.4+284038 & 0.5468 & 0304071301 & 480 & N & Y & Y & N & Y\\
\hline
3XMMJ093458.3+611233 & 0.245 & 0085640201 & 391 & N & Y & N & Y & Y\\
\hline
3XMMJ093532.9+612739 & 0.4755 & 0085640201 & 112 & N & Y & N & Y & Y\\
\hline
3XMMJ093551.5+612111 & 0.0393 & 0085640201 & 1028 & Y & Y & N & Y & Y\\
\hline
3XMMJ093952.7+355358 & 0.1367 & 0021740101 & 1195 & Y & Y & Y & N & Y\\
\hline
3XMMJ095902.7+021906 & 0.3454 & 0203362301 & 1046 & N & N & N & N & N\\
 &  & 0203362201 & 2459 & N & N & N & N & N\\
 &  & 0203361801 & 2851 & N & N & N & N & N\\
 &  & 0302351801 & 1088 & N & N & N & N & N\\
 &  & 0501170201 & 1208 & N & N & Y & N & Y\\
 &  & 0302351701 & 688 & N & N & N & N & N\\
 &  & 0302352301 & 110 & N & N & N & N & N\\
 &  & 0302352201 & 282 & N & N & N & N & N\\
 &  & 0203361701 & 1150 & N & N & N & N & N\\
\hline
3XMMJ095906.6+130134 & 0.0366 & 0504100201 & 1937 & N & Y & Y & N & Y\\
\hline
3XMMJ100032.2+553630 & 0.215 & 0110930201 & 246 & N & N & Y & N & Y\\
 &  & 0147760101 & 227 & N & Y & N & N & Y\\
\hline
3XMMJ100129.4+013633 & 0.1042 & 0203360501 & 152 & N & Y & N & Y & Y\\
 &  & 0302351001 & 553 & N & Y & N & N & Y\\
 &  & 0203361001 & 516 & N & Y & N & N & Y\\
\hline
3XMMJ101830.8+000504 & 0.0623 & 0402781401 & 1587 & N & Y & N & N & Y\\
 &  & 0402780101 & 668 & N & Y & N & N & Y\\
\hline
3XMMJ103408.5+600152 & 0.051 & 0306050701 & 745 & Y & Y & Y & N & Y\\
\hline
3XMMJ104451.7+063548 & 0.0276 & 0405240901 & 1275 & Y & N & N & N & Y\\
\hline
3XMMJ104522.1+212614 & 0.8908 & 0128531601 & 4732 & N & N & N & N & N\\
 &  & 0128531501 & 921 & N & N & N & N & N\\
 &  & 0128531401 & 571 & N & Y & N & Y & Y\\
\hline
3XMMJ104930.9+225752 & 0.0327 & 0312191501 & 1260 & N & Y & Y & N & Y\\
\hline
3XMMJ111217.9+132106 & 0.2315 & 0500760101 & 299 & N & N & Y & N & Y\\
\hline
3XMMJ112026.6+431518 & 0.1459 & 0107860201 & 184 & N & Y & N & Y & Y\\
\hline
3XMMJ112345.1+061605 & 0.2338 & 0103863201 & 269 & N & N & Y & N & Y\\
\hline
3XMMJ113240.2+525701 & 0.0266 & 0200431301 & 620 & N & Y & Y & N & Y\\
 &  & 0200430501 & 919 & N & Y & Y & N & Y\\
\hline
3XMMJ113409.0+491516 & 0.0372 & 0149900201 & 1170 & N & Y & N & N & Y\\
\hline
3XMMJ113549.0+565708 & 0.0511 & 0504101001 & 697 & N & Y & Y & N & Y\\
\hline
3XMMJ114209.9+600435 & 0.3477 & 0502780301 & 988 & Y & N & N & N & Y\\
\hline
3XMMJ114454.8+194635 & 0.0274 & 0061740101 & 196 & N & Y & N & Y & Y\\
\hline
3XMMJ115704.8+524903 & 0.0355 & 0504100901 & 885 & Y & Y & Y & N & Y\\
\hline
3XMMJ120349.1+020556 & 0.081 & 0093060101 & 371 & N & Y & N & Y & Y\\
\hline
3XMMJ120429.6+201858 & 0.0226 & 0112271001 & 260 & N & Y & Y & N & Y\\
 &  & 0112271101 & 381 & N & Y & Y & N & Y\\
 &  & 0112270601 & 555 & N & Y & N & N & Y\\
\hline
3XMMJ120629.3+281435 & 0.2927 & 0301900401 & 99 & N & Y & N & Y & Y\\
\hline
3XMMJ121118.8+503652 & 0.1023 & 0203170101 & 4145 & N & Y & N & Y & Y\\
\hline
3XMMJ121839.4+470626 & 0.0939 & 0203270201 & 188 & N & Y & Y & N & Y\\
 &  & 0400560301 & 158 & N & Y & Y & N & Y\\
\hline
3XMMJ121923.2+054929 & 0.0073 & 0502120101 & 46994 & Y & Y & N & N & Y\\
 &  & 0056340101 & 16369 & N & Y & N & N & Y\\
\hline
3XMMJ122135.6+280614 & 3.2877 & 0502211101 & 203 & N & N & N & N & N\\
 &  & 0502211301 & 465 & N & N & N & N & N\\
 &  & 0502211401 & 168 & N & N & N & N & N\\
 &  & 0104860501 & 4615 & N & Y & N & Y & Y\\
 &  & 0502211201 & 137 & N & N & N & N & N\\
\hline
3XMMJ122349.5+072657 & 0.0013 & 0205090101 & 1130 & Y & N & N & N & Y\\
\hline
3XMMJ122546.7+123943 & 0.0086 & 0110930301 & 5250 & N & Y & N & Y & Y\\
 &  & 0110930701 & 19165 & N & Y & Y & N & Y\\
 &  & 0675140101 & 116555 & Y & Y & N & N & Y\\
\hline
3XMMJ122548.8+333248 & 0.001 & 0112522001 & 2062 & N & Y & Y & N & Y\\
 &  & 0142830101 & 145167 & N & Y & N & Y & Y\\
 &  & 0112522701 & 11519 & N & Y & N & Y & Y\\
 &  & 0112521901 & 10640 & N & Y & Y & N & Y\\
\hline
3XMMJ123056.2+155212 & 0.1877 & 0106061001 & 200 & N & Y & N & Y & Y\\
\hline
3XMMJ123625.4+125843 & 0.0934 & 0200650101 & 1313 & N & Y & N & N & Y\\
\hline
3XMMJ123719.3+114915 & 0.1075 & 0112840101 & 101 & N & Y & N & Y & Y\\
\hline
3XMMJ123843.4+092736 & 0.0829 & 0504100601 & 2876 & N & Y & Y & N & Y\\
\hline
3XMMJ125710.7+272417 & 0.0207 & 0652310401 & 1060 & N & N & N & N & N\\
 &  & 0652310501 & 79 & N & N & N & N & N\\
 &  & 0652310301 & 295 & N & N & N & N & N\\
 &  & 0652310601 & 61 & N & N & N & N & N\\
 &  & 0403150201 & 606 & N & N & N & Y & Y\\
 &  & 0652310701 & 986 & N & N & N & N & N\\
 &  & 0124710101 & 1277 & N & N & N & N & N\\
 &  & 0652311001 & 744 & Y & N & N & N & Y\\
 &  & 0403150101 & 288 & N & Y & N & Y & Y\\
 &  & 0652310901 & 1763 & N & N & N & N & N\\
 &  & 0652310201 & 724 & N & N & N & N & N\\
 &  & 0652310801 & 714 & N & N & N & N & N\\
\hline
3XMMJ125724.3+272952 & 0.0245 & 0124712201 & 213 & N & N & N & N & N\\
 &  & 0652310601 & 187 & N & N & N & N & N\\
 &  & 0652311001 & 156 & N & N & N & N & N\\
 &  & 0652310901 & 509 & N & N & N & N & N\\
 &  & 0652310801 & 448 & N & N & N & N & N\\
 &  & 0652310201 & 277 & N & N & N & N & N\\
 &  & 0652310701 & 449 & N & N & N & N & N\\
 &  & 0124710101 & 991 & Y & N & N & N & Y\\
 &  & 0403150201 & 1288 & N & N & N & N & N\\
 &  & 0403150101 & 978 & N & N & N & N & N\\
 &  & 0652310401 & 769 & N & N & N & N & N\\
\hline
3XMMJ125930.9+282705 & 1.0939 & 0204040301 & 109 & N & N & N & N & N\\
 & & 0204040201 & 216 & N & Y & N & N & Y\\
 & & 0204040101 & 514 & N & Y & N & Y & Y\\
 & & 0304320201 & 154 & N & N & N & Y & Y\\
\hline
3XMMJ130128.1+275106 & 0.2425 & 0124710901 & 886 & Y & N & N & N & Y\\
 &  & 0124710801 & 1308 & N & N & N & N & N\\
\hline
3XMMJ131104.6+272806 & 0.2397 & 0021740201 & 423 & N & Y & N & N & Y\\
\hline
3XMMJ132348.4+431804 & 0.0273 & 0504101601 & 1149 & N & Y & Y & N & Y\\
\hline
3XMMJ132410.0+135835 & 0.023 & 0108860701 & 3164 & Y & N & N & N & Y\\
\hline
3XMMJ133807.5+280508 & 1.0881 & 0110940101 & 567 & N & Y & N & Y & Y\\
\hline
3XMMJ134245.8+403913 & 0.0893 & 0070340701 & 857 & N & Y & Y & N & Y\\
\hline
3XMMJ134459.4-001559 & 0.2449 & 0111282501 & 507 & N & N & Y & N & Y\\
 &  & 0111281801 & 625 & N & N & N & N & N\\
\hline
3XMMJ134507.9-001900 & 0.4189 & 0111282501 & 363 & N & Y & N & N & Y\\
\hline
3XMMJ134656.6+580316 & 0.3722 & 0112250201 & 188 & N & Y & N & Y & Y\\
\hline
3XMMJ134745.6+264053 & 0.2529 & 0205190101 & 460 & N & N & Y & N & Y\\
\hline
3XMMJ134834.9+263109 & 0.0589 & 0109070201 & 20801 & Y & N & N & N & Y\\
 &  & 0097820101 & 49559 & N & N & N & N & N\\
\hline
3XMMJ135436.3+051524 & 0.0815 & 0404240101 & 272 & N & Y & N & Y & Y\\
\hline
3XMMJ140040.5-015518 & 0.025 & 0200430901 & 132 & N & N & N & N & N\\
 &  & 0505930101 & 110 & N & N & N & N & N\\
 &  & 0505930401 & 115 & N & N & Y & Y & Y\\
\hline
3XMMJ140700.3+282714 & 0.0766 & 0140960101 & 1110 & Y & Y & N & Y & Y\\
\hline
3XMMJ141546.2+112943 & 2.5603 & 0112250301 & 156 & N & Y & N & Y & Y\\
 &  & 0112251301 & 163 & N & Y & N & Y & Y\\
\hline
3XMMJ141602.1+360923 & 0.171 & 0148620101 & 569 & N & Y & N & N & Y\\
\hline
3XMMJ143025.8+415956 & 0.3524 & 0212480701 & 430 & N & Y & N & Y & Y\\
 &  & 0111260101 & 345 & N & N & N & N & N\\
 &  & 0111260701 & 1458 & N & N & N & N & N\\
\hline
3XMMJ144659.9+025330 & 0.0013 & 0203050401 & 1072 & N & Y & N & Y & Y\\
 &  & 0203050801 & 3488 & N & Y & N & Y & Y\\
\hline
3XMMJ145301.4+164452 & 0.8386 & 0091140201 & 352 & N & Y & N & Y & Y\\
\hline
3XMMJ145442.2+182937 & 0.1163 & 0145020101 & 552 & N & Y & N & N & Y\\
\hline
3XMMJ145720.4-011102 & 0.0873 & 0502780601 & 75 & N & Y & N & Y & Y\\
\hline
3XMMJ150121.9+014401 & 0.6082 & 0554680201 & 2432 & N & Y & N & N & Y\\
 &  & 0554680301 & 1848 & N & Y & N & Y & Y\\
 &  & 0302460101 & 731 & N & Y & N & N & Y\\
\hline
3XMMJ150754.3+010817 & 0.061 & 0402781001 & 394 & N & Y & Y & N & Y\\
\hline
3XMMJ150946.8+570002 & 0.0026 & 0111260201 & 109 & N & Y & Y & N & Y\\
\hline
3XMMJ151106.4+054123 & 0.0807 & 0111270201 & 2746 & N & N & N & N & N\\
 &  & 0551780301 & 3743 & N & Y & N & Y & Y\\
 &  & 0551780201 & 4558 & N & Y & N & Y & Y\\
 &  & 0551780501 & 1949 & N & Y & N & Y & Y\\
 &  & 0551780401 & 4369 & N & Y & N & Y & Y\\
\hline
3XMMJ153457.2+233011 & 0.0181 & 0101640901 & 739 & Y & Y & N & N & Y\\
 &  & 0205510501 & 502 & Y & N & N & N & Y\\
 &  & 0205510201 & 340 & N & Y & N & N & Y\\
 &  & 0101640801 & 685 & Y & N & N & N & Y\\
 &  & 0205510401 & 176 & N & N & N & N & N\\
\hline
3XMMJ153641.6+543505 & 0.447 & 0150610301 & 111 & N & Y & N & Y & Y\\
 &  & 0300310401 & 374 & N & Y & N & N & Y\\
 &  & 0300310501 & 491 & N & Y & N & N & Y\\
\hline
3XMMJ160426.5+174431 & 0.041 & 0147210301 & 375 & N & Y & N & N & Y\\
\hline
3XMMJ205017.8-053626 & 6.0E-4 & 0203050501 & 11300 & N & Y & N & Y & Y\\
\hline
3XMMJ215649.5-074531 & 0.0554 & 0654440101 & 256 & N & Y & N & N & Y\\
 &  & 0404910701 & 98 & N & Y & Y & N & Y\\
\hline
\hline
\end{longtable}
\begin{list}{}{}
\item
{\it Columns:}(1) Source name in the 3XMM-DR4 catalog. (2)
  Spectroscopic redshift from SDSS-DR7. (3) Observation identifier in
  3XMM-DR4. (4) Counts collected (all instruments added together) in
  0.5-10 keV. (5), (6), (7), and (8): whether or not the automated
  spectral-fitting results flagged the observation as belonging to
  each sample. (9) Whether or not the observation fulfill any of the
  samples criteria.
\end{list}
\end{landscape}
}
\end{document}